\documentclass[]{aa}
\usepackage{times}
\usepackage{graphics}
\usepackage{epsf}
\begin{document}

\title{Efficient Monte Carlo methods for continuum radiative transfer}

\titlerunning{Efficient Monte Carlo methods}

\author{
M.~Juvela 
}

\offprints{M.~Juvela}

\institute{
Helsinki University Observatory, T\"ahtitorninm\"aki, P.O.Box 14,
SF-00014 University of Helsinki, Finland (mjuvela@astro.helsinki.fi) \\
}  

\date{Received <date> ; accepted <date>}

\abstract{ 
We discuss the efficiency of Monte Carlo methods in solving
continuum radiative transfer problems. The sampling of the radiation field and
convergence of dust temperature calculations in the case of optically thick
clouds are both studied. 
For spherically symmetric clouds we find that the computational cost of Monte
Carlo simulations can be reduced, in some cases by orders of magnitude, with
simple importance weighting schemes. This is particularly true for models
consisting of cells of different sizes for which the run times would otherwise
be determined by the size of the smallest cell. We present a new idea of
extending importance weighting to scattered photons. This is found to be
useful in calculations of scattered flux and could be important for
three-dimensional models when observed intensity is needed only for one
general direction of observations.
%
Convergence of dust temperature calculations is studied for models with
optical depths $\tau_{\rm V}=10-10^4$.
We examine acceleration methods where 
radiative interactions inside a cell or between neighbouring cells are treated
explicitly. In optically thick clouds with strong self-coupling between dust
temperatures the run times can be reduced by more than one order of magnitude.
The use of a reference field was also examined. This eliminates the need for
repeating simulation of constant sources (e.g., background radiation) after the
first iteration and significantly reduces sampling errors. The applicability
of the methods for three-dimensional models is discussed. 
 }

\maketitle

\section{Introduction}

Continuum radiative transfer problems are commonly solved with Monte Carlo
calculations. The radiation field is simulated according to the actual
processes that are assumed to be taking place in an interstellar cloud.
Emission from background, internal sources, and the dust itself is represented by a
number of photons packages, each corresponding to a large number of actual
photons at one wavelength. Random numbers determine the initial positions
and directions of the photon packages.  The distance to a point where
scattering takes place and the direction after scattering are also calculated
using random numbers. This is the basic scheme in studies of light
scattering (e.g., Mattila \cite{mattila70}; Witt \cite{witt77}). In scattering,
the probability distribution for the change of photon direction is given by
the scattering function that depends on dust properties. It is this random
change of direction that precludes (with the exception of isotropic scattering
or pure forward scattering) the use of direct ray tracing methods which are
often used in line transfer. If, in addition to scattered light, also dust
emission is to be solved then during the simulation the number of absorbed
photons must be registered in each cell of the model cloud. This could be done
at each position where scattering takes place but unless optical depth is very
high (or one is calculating only the scattered flux) it is better to
explicitly calculate absorptions in all cells that the photon package passes
through.  Especially in optically thin clouds the statistics of absorbed
energy would otherwise remain very poor. 

Monte Carlo continues to be the standard method for the modelling of dust
emission from interstellar clouds (Bernard et al. \cite{bernard92}; for recent
papers see, e.g., Stamatellos \& Whitworth \cite{stamatellos}; Concalves et al.
\cite{concalves}; Niccolini et al.\cite{niccolini}; Juvela \& Padoan
\cite{juvela03}; Pascucci et al. \cite{pascucci} and references therein; 
Kurosawa et al. \cite{kurosawa04}; Whitney et al. \cite{whitney03}; Wolf et
al. \cite{wolf03}). Several methods can be employed to improve the sampling.
One of the most common is the method of forced first scattering (e.g., Mattila
\cite{mattila70}) which improves sampling of scattered flux in clouds of low
optical depth. In an optically thin cloud most photons pass through the cloud
unscattered and give no information on scattered flux. Forced scattering means
that one calculates for each package the fraction of photons that do scatter
in the cloud. Unscattered photons are followed through the cloud along the
original direction while the calculated fraction is always scattered somewhere
along the way. The distance to the point of scattering is calculated from a
conditional probability distribution 'where will the photon package scatter if
it does scatter before exiting the cloud'.

In optically thick clouds most of the incoming flux is scattered many times
and a photon package does not usually propagate very far.  In a spherically
symmetric model this means that very few background photon packages reach the
cloud centre. In order to get proper sampling in optically thick regions the
number of simulated photon packages must be extremely large, and the run times
become correspondingly very long.  Niccolini et al. (\cite{niccolini}; method
2) presented a partial solution which is analogous to the method of forced
scattering but where the roles of the scattered and unscattered flux are
reversed. The fraction of photons that does {\em not} scatter is first
calculated and that part of the photon package is followed through the cloud
toward the original direction. Rest of the photons is scattered somewhere
along the way. The procedure is repeated after each scattering.  While the
method of forced scattering gives (in statistical sense) the right amount of
scattered flux it is not equally clear that the method of Niccolini et al.
(\cite{niccolini}) always provides correct estimates of absorbed photons. If
the number of simulated photon packages is small the flux observed in cloud
centre consist only of photons that are first scattered a few times near the
surface and after this propagate unscattered to the centre. It remains open
what is the ratio of these photons and those in the more rare photon packages
(not necessarily more rare photons) that reach the centre only after many
scatterings.  At the limit of high optical depth for scattering and
low optical depth for absorption the method would {\em appear} to improve the
sampling but actually all flux in the centre should still result from packages
that have scattered numerous times, i.e. the method could not decrease the
number of photon packages needed in the simulation. These are, of course, not
faults of the method itself but are simply consequences of the sampling
problem. The method does not address problems arising from different
cell sizes and, therefore, while this method is probably very
useful in many cases it is not yet a complete solution to the problem.

One of the main advantages of Monte Carlo methods is their great flexibility
in the way radiation field is sampled. All probability distributions
employed in the simulation of photon packages can be modified as long as
these are compensated by corresponding changes in the weighting, i.e. the
number of actual photons in a package. For some reason this advantage is
often not used and simulations are unnecessarily inefficient. High optical
depths are perceived to pose another problem and not only because of
difficulties in the sampling of the radiation field. In an optically thick
cloud there may be significant self-coupling between dust temperatures. In
the usual scheme this means that dust temperatures can not be solved
directly and one is forced to iterate between simulation of the radiation
field and updating of dust temperature estimates. The number of iterations
depends on the optical depth, $\tau$. For a cloud where $\tau$ reaches at
optical wavelengths several hundred run times can become orders of magnitude
longer than in the case of an optically thin cloud. This depends, however,
very much on the external radiation field and the dust temperatures.
According to Bernard et al. (\cite{bernard92}), for cold clouds embedded in
normal interstellar radiation field the dust coupling is unimportant unless
optical depth is several hundred. This was confirmed by Concalves et al.
(\cite{concalves}; see their Fig. 1) but the situation can be very different
in the presence of hot dust. 

The slow convergence at high optical depths is, of course, not related to the
Monte Carlo method per ce. Even if radiation field is sampled with Monte
Carlo method the dust temperatures {\em can} be solved without iterations and
even if iterations are made, simulation of the radiation field need not be
repeated. Because of the larger memory requirements such schemes are,
however, practical only for one-dimensional models. 
Bjorkman \& Wood (\cite{bjorkman01}) presented a modification where local dust
temperature is updated after each absorption and a new photon is immediately
re-emitted from the same position. The frequency of the emitted photon is
obtained from a probability distribution that takes into account both the
previously emitted photons and the correction due to the updated temperature.
The desired noise level determines the number of required photon packages and
once these and the induced re-emitted photons have been simulated the
calculations provide a converged solution. The authors state that the method
does not suffer from convergence problems associated with $\Lambda$-iteration
methods. We return to this question later in the paper.

In this paper we will study the efficiency of Monte Carlo radiative transfer
calculation. We will discuss separately efficient sampling of the radiation
field and convergence of iterations. One-dimensional, i.e. spherically
symmetric model clouds are used as examples. We will, however, emphasise that
most results can be directly transferred to three-dimensional cloud modelling
and we will briefly discuss the applicability of the methods in the case of
3D models containing millions of cells. In Sect.~\ref{sect:program} we describe
the program and the implementation of weighting, reference field and
convergence acceleration schemes.  Results from tests with one-dimensional
models are presented in Sect.~\ref{sect:1d}, and the results are discussed in
Sect.~\ref{sect:discussion}

\section{The program} \label{sect:program}

In this section we present some details of our implementation of the
radiative transfer program. More specifically, we discuss the possibilities
of improving the sampling of the radiation field so that a given accuracy of
the results can be obtained with a smaller number of photons packages. On
the other hand, we will discuss the convergence of dust temperature
calculations in the case of optically thick clouds. This part of the
discussion does apply equally to programs where some method other than Monte
Carlo is used to estimate the radiation field.

\subsection{Sampling of the radiation field} \label{sect:sampling}

One of the main advantages of Monte Carlo methods is that the sampling can be
easily adapted according to the problem at hand. This possibility is, however,
seldom used and, in order to ensure proper sampling in all parts of the model,
simulations need an exceedingly large number of photon packages. This
adaptation or `weighting' is what in Monte Carlo integration would be called
importance sampling. While it may be possible to create adaptive methods
(analogous to stratified sampling; see Press et al. \cite{press_nr}) we
restrict our discussion to more simple schemes.

In normal Monte Carlo simulation photon packages are created at random
locations (within emitting medium or on the surfaces of separate
radiation sources) and sent uniformly toward random directions. Each package
has equal weight i.e. the number of true photons is divided evenly between
created packages. One is not restricted to the use of uniform spatial and
angular distributions and one could, e.g., send twice as many photon packages
from certain part of the cloud provided that each of these packages contains
only half of the original number of photons. More generally, if one assumes a
new probability distribution $p(r,\Omega)$ for positions $r$ and directions
$\Omega$ of the created photon packages the weight of the photon package is
\begin{equation}
W_i = \frac{1}{p(r_i,\Omega_i)}.
\end{equation}
The default number of photons included in a package is
multiplied with this number where $r_i$ and $\Omega_i$ are the actual position
and direction of the photon package. Weighting can be applied to background
photons, photons emitted by dust inside the cloud and for each discrete source
included in the model. In each case the spatial and angular distributions can
be weighted separately. 
The four weighting schemes used in this paper ($A$, $B$, $C$, and $D$)
are illustrated in Fig.~\ref{fig:schemes}. These are are discussed below
and a more detailed presentation can be found in Appendix~\ref{sect:app_weighting}.

\begin{figure}
\resizebox{\hsize}{!}{\includegraphics{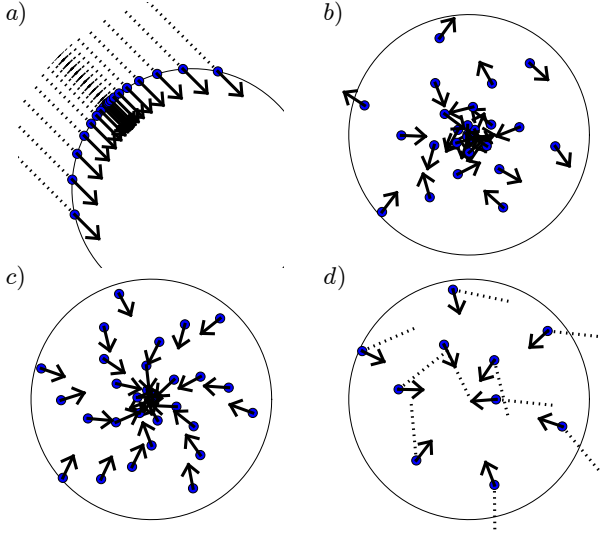}} 
\caption[]{
%
The four weighting schemes ($A$, $B$, $C$, and $D$) used in the Monte Carlo
simulations in this paper. The figures illustrate a case where the sampling is
improved in the centre of a spherically symmetric model cloud: background
packages are sent preferentially towards the cloud centre (scheme $A$), more
photon packages are created in the inner parts (scheme $B$), within the cloud
photon packages are sent preferentially towards the cloud centre (scheme $C$),
and scattered photon packages are directed preferentially towards the cloud
centre (scheme $D$).

%
}
\label{fig:schemes}
\end{figure}

\subsubsection{External radiation field: weighting scheme $A$}


%
In a spherically symmetric cloud all background photon packages enter the
outermost shell but, because of their smaller diameter, the inner shells are
hit by relatively few packages. At least in optically thin clouds the problem
can be alleviated by sending photon packages preferentially toward cloud
centre. The original probability distribution of the impact parameter, $p(d)
\propto d$, can be replaced one with higher probability at small values of $d$
(e.g., $p(d) \propto d^{\alpha}$ with $\alpha<1$). We use a scheme where the
same number of photon packages is sent towards each annulus as defined by the
radial discretization (weighting scheme $A$). Within each annulus we have the
usual distribution of $p(d) \propto d$.  The number of photon packages is
fixed for each annulus and the random noise is reduced. This is, however, not
yet optimal. In an optically thin model with $N$ shells each photon package
enters the outermost shell but only one in $N$ hits the innermost cell. A
more strongly peaked probability distribution could result in more uniform
errors.
%

\subsubsection{Internal radiation field: weighting scheme $B$}


%
Different cell sizes also affect the sampling of dust emission within the
cloud. This is particularly problematic in optically thick clouds where
self-coupling between dust temperatures is strong. In a spherical model the
the innermost cells can be orders of magnitude smaller than the the outermost
shells. Sampling errors increase toward centre and the number of required
photon packages depends directly on the size of the smallest cells. The
solution is similar as in the case of background photons.  Emissions take
place at random positions with radial probability distribution, $p(r)\sim
r^2$. For photon packages this can be replaced with a steeper distribution
(e.g., $p(r)\sim r^{\beta}$ with $\beta$ below 2). We use a procedure where the
same number of photon packages is generated in each cell but the locations
within the cells follow the $p(r)\sim r^2$ distribution (weighting scheme
$B$).  In extremely optically thick clouds one might also consider whether
photon packages should be created preferentially close to cell boundaries so
that these would better sample the energy transfer between cells. The
importance of different cells as sources of radiation could also be taken into
account and more packages could be sent from denser and warmer cells.
%

\subsubsection{Angular distribution of photons created in the cloud: weighting
scheme $C$}


%
In order to improve the sampling of the smallest cells we consider
weighting the angular distribution of photon packages created within the
cloud. Uniform angular distribution is replaced with distribution that peaks
in the direction of the cloud centre or, more generally, toward any region
where sampling is to be improved. We use an exponential function
$p(\theta)\propto e^{-\gamma \theta}$ where $\theta$ is angular offset from
the direction of the cloud centre and $\gamma$ is a positive constant
(weighting scheme $C$). The exponential function is convenient since the
inverse cumulative probability density function, $P^{-1}$ , is easily
calculated and can be used to generate angles from the selected distribution.
However, the calculation of the $P^{-1}$ can be replaced with a lookup to a
table of $P$-values and, in practise, there are no restrictions for using any
function as the probability distribution $p$. 
%

\subsubsection{Place and angular distribution of scattering: weighting scheme $D$}

In weighting scheme $D$ we apply weighting to scattered photons. The method of
forced first scattering would be one example of this (see
Sect.~\ref{sect:app_basic}). Here we are, however, more interested in the
angular distribution of the scattered photons. The goal can be to increase the
flux of packages (not photons!) toward optically thick parts of the cloud or
toward one particular direction from which the emerging scattered intensity is
observed. Scattered photons already have a non-uniform probability
distribution, $p(\Omega)$, which is determined by scattering function of the
dust model. This can be given by the Henyey-Greenstein (\cite{henyey41})
formula, or some other function specified in analytical or tabular form. The
distribution $p$ is defined in a coordinate system that is fixed by the
initial direction of the photon package. Let us assume that we want to have
locally a probability distribution $q(\Omega)$ for the directions of scattered
photon packages. Random directions are generated according to this distribution and
each package is weighted with the ratio
\begin{equation}
   W_i =   \frac{p(\Omega')}{q(\Omega)}. \label{eq:weight}
\end{equation}
Functions $p$ and $q$ are defined in different coordinate systems and one must
apply a rotation to find out what direction $\Omega'$ the selected
direction $\Omega$ corresponds to.

The presented scheme clearly works if one divides a photon package into $n$
small packages, each containing a fraction $\propto p(\Omega')/q(\Omega)$ of
the original number of photons and having directions following the
distribution $q$. 
%
%
For practical reasons it is not possible to split a package this way. One can,
however, use the scheme without creating any new photon packages. When a
package is scattered the new direction is obtained from the probability
distribution $q$ and the number of photons contained in the package is
multiplied by the factor given by Eq.~\ref{eq:weight}. The weight $W_i$ can be
smaller or larger than one and for some photon packages the number of
photons can actually increase. The result is, of course, correct only
statistically and after the simulation of many photon packages.  The
possibility of weighting leads to an interesting conclusion that all
frequencies can be calculated simultaneously using a single photon package.
Each time scattering occurs the number of photons at each frequency is
multiplied by a factor depending on the realised scattering angle and the
scattering function at that frequency. Weighting should also be applied
according to the distance between scatterings which depends on the
frequency-dependent opacity of the medium. Since this is very wavelength
dependent such a scheme would, however, most likely result in increased
sampling errors.
%

The method $D$ is not the same as the 'peeling off' method of Yusef-Zadeh et
al. (\cite{yusef94}) that is used to improve the quality of images of
scattered light (Wood \& Reynolds \cite{wood99}; Whitney et al.
\cite{whitney03}). In the 'peeling off' method one calculates and registers
after each scattering the fraction of photons that are scattered towards the
observed and escape the cloud. The angular distribution (or weighting) of the
rest of the photons still essentially follows the original scattering
function. Therefore, the sampling of the scattered flux remains unchanged
within the cloud while in our method $D$ it is changed. Method $D$ improves the
sampling of the observed scattered flux by increasing the number of regular
photon packages that exit the cloud towards the observer. The 'peeling off'
method is more efficient if only emerging flux is considered and only for
one single direction. Method $D$ might be competitive in cases where observed
flux results from multiple scatterings in an optically thick medium. Because
method $D$ improves sampling in the selected general direction it can also be
used when images of the scattered flux are needed for several directions close
to each other. Of course, it would be possible to combine the two methods.
Method $D$ would be used to modify sampling within the cloud and the 'peeling
off' could still be performed after each scattering. This possibility is not
studied any further in this paper. 

\subsection{Reference field}

In Monte Carlo integration errors are proportional to function values. If
solution is known for a reference function that is similar to the integrated
function only the difference needs to be solved with Monte Carlo methods and
the errors are proportional to the difference between the actual function and
the reference. This idea was first applied to line transfer calculations by
Bernes (\cite{bernes79}). He assumed a reference field corresponding to a
fixed excitation temperature and Monte Carlo simulation was used to determine
the differences between the true field and the constant reference field. Choi
et al. (\cite{choi}) improved the method by using the solution from the
previous iteration as the reference. As the reference field approaches the
true field the sampling errors decrease on each iteration. 

%
So far a reference field has not been used in connection with dust emission
calculations. The implementation is, however, straightforward. The
reference field is taken to correspond to the situation of the previous
iteration. On the first iteration reference field is zero and calculations
proceed in the normal fashion. On the following iterations each created photon
package contains the difference between the true number of photons
(corresponding to the latest temperature estimates) and the photons from the
reference field (corresponding to the previous temperature estimates). For
constant radiation sources and the background radiation this difference is
zero and simulation is needed only to find out the effect of the latest dust
temperature updates. 
%


%
The concept of a reference field is useful in calculations involving many
iterations. First of all, emission from constant sources (e.g., background)
can be simulated on a single iteration. On the following iterations their
effect would already be included in the reference field and no further photon
packages need to be simulated from them. Secondly, sampling errors are
decreased (or a given accuracy is reached with fewer photon packages) since
the final noise level depends on the total number of simulated photon packages
rather than the number of packages per one iteration. Details of the
implementation are given in Appendix~\ref{sect:app_reference}. In principle,
the use of a reference field does not affect the convergence of dust
temperatures that is discussed below.
%

\subsection{Accelerated iterations} \label{sect:accel}

Only in the optically thin case ($\tau<<1$) can dust temperatures be solved
directly. In optically thick clouds the dust emission can contribute
significantly to dust heating and this leads to iterations where radiation
field and dust temperatures are solved alternatingly.  As $\tau$ increases,
more and more iterations are needed and calculations can potentially become
very time-consuming. 

Dust temperatures can, in principle, be computed without repeating the
simulation of the radiation field. Once the radiative coupling between all
cells has been calculated one can write a set of equations from which dust
temperature in all cells can be solved. The same applies to, e.g., molecular
line calculations where the unknowns are level populations in different cells.
As noted by  Stamatellos et al. (\cite{stamatellos}), the case is much
simpler for the continuum radiative transfer in dust clouds. Absorption and
scattering cross sections can be assumed to be independent of temperature and
radiative coupling (e.g., 'the fraction of photons emitted from cell $i$ is
absorbed in cell $j$') remains constant. If information about the coupling is
saved a simulation providing this information needs to be done only once. If
one considers dust particles at equilibrium temperature the whole problem
reduces to a set of non-linear equations. The problem becomes more complicated
if one must consider in each cell a distribution of dust temperatures.
Temperature distribution is usually solved separately from a linear set of
equations where the unknowns (i.e. number of dust particles in a given
temperature interval) are multiplied with factors that depend on temperature
distributions in other cells. The whole problem (including radiative
couplings) can still be formulated as a single set of non-linear equations.

For one-dimensional problems where only dust particles at an equilibrium
temperature are considered the solution of such equations is still feasible.
For transiently heated particles or for any models in several dimensions this
is no longer practical. For a model of $N$ cells one would have to store for
each simulated frequency $(N^2-N)/2$ terms describing the coupling between any
two cells. If these can not be stored then also the simulation of the
radiation field needs to be repeated each time before temperatures can be
updated. For optically thin clouds iterations are never needed. Once the
optical depth becomes high hundreds of iterations may be required and it
becomes difficult to follow the convergence if results include random sampling
errors. The slowness of convergence is, of course, not caused by the Monte
Carlo simulation which is only a method for estimating the radiation field.

As $\tau$ increases most emitted photons are absorbed locally and a larger
fraction of energy flow takes place within a cell with relatively small amount
of interaction between cells. In calculations this translates to small
temperature corrections and slow convergence of dust temperatures.  The number
of iterations needed to reach correct dust temperatures depends not only on
optical depth but also on dust temperatures and the spectrum of the external
radiation field. If dust is cold and, in spite of high visual optical depth,
cloud is penetrated with sufficient amount of longer wavelength external
radiation the self coupling of dust temperatures may remain unimportant. 

There are several ways to improve the convergence rate. We will consider two
alternatives, a heuristic extrapolation based on dust temperatures adopted
from previous iterations, and the accelerated Monte Carlo methods. Accelerated
Monte Carlo schemes (AMC) are analogous to ALI (Accelerated Lambda Iteration)
methods and in these part of radiative interactions are treated explicitly
when dust temperatures are updated.  Such methods have already been used in
line transfer (Juvela \& Padoan \cite{juvela01}; Hogerheijde \& van der Tak
\cite{hogerheijde}). We will concentrate on the case where we can assume that
dust is in each cell at one equilibrium temperature. The generalisation to the
case of a dust temperature distribution is rather straightforward.

\subsubsection{Heuristic extrapolation} \label{sect:accelxpol}

Extrapolation is based on dust temperatures at three consecutive iterations.
We denote the latest change per iteration $(\Delta T/iter)_i$, and the factor
$k$ by which this derivative was seen to change, i.e. $(\Delta T/iter)_i = k
(\Delta T/iter)_{i-1}$. If $0<|k|<1$, i.e. the temperatures show signs of
convergence, an extrapolation is done. This assumes that each successive
temperature correction will continue to be $k$ times the previous one. The
extrapolation approaches a constant value but in our implementation we make
extrapolation only 20 iterations forward. The temperature correction obtained
from the extrapolation is further limited to maximum of 20\% of the original
value. The method requires some extra memory (two values per cell) but very
little computations. In the simulation the same set of random numbers should
be used on each iteration so that random variations do not interfere with the
extrapolation.

\subsubsection{AMC methods} \label{sect:accelali}

In normal calculations one must compute in each cell the strength of the
radiation field which consists of the intensity produced by the cell itself
and the intensity caused by every other cell. Later this information is used
to compute new estimates of dust temperatures. The other extreme case was
mentioned above: the coupling between cells is determined and temperatures are
computed simultaneously from a large set of equations. In between there are
many alternatives where part of the interactions are treated explicitly in
dust temperature calculations. This is analogous to Accelerated Lambda Methods
(ALI) used in line transfer (Rybicki \& Hummer \cite{rh91}, \cite{rh92}).
There, the local intensity $J$ is computed from source functions $S$ through
lambda operator, $J=\Lambda S$, and the operator is split into two parts. In
the most common case the diagonal part of the operator is separated. This
represents the intensity produced in a cell by the cell itself. In an
optically thick case most of the created photons are absorbed locally. This
loop does not contribute to a change in level populations but its elimination
does significantly improve the rate of convergence. 

In Monte Carlo we calculate photon absorptions rather than the intensity. The
use of diagonal $\lambda$ operator is replicated by calculating in each cell
only those absorbed photons that were emitted outside that cell. At the same
time we note for each cell the photon escape probability i.e. the fraction of
photons that escapes the emitting cell. One must remember to take into account
those photons that first escape the cell but are later scattered back and
absorbed. In dust temperature calculations absorbed photons originating from
other cells, discrete radiation sources and from the background are balanced
against that fraction of emitted photons that leave the emitting cell.
Temperature calculations do not become more complicated but some additional
memory is required as the photon escape probabilities must be saved for each
cell and each simulated frequency. In spite of this the method is feasible
even for three-dimensional models. We will call this the AMC-method with
diagonal operator.

We have also implemented a `tridiagonal operator'. For one-dimensional models
this means that when dust temperatures are updated the radiative coupling
between closest neighbours is treated explicitly. In simulations we ignore
those absorbed photons that were emitted by the cell itself or one of its
immediate neighbours. Additionally, we note again the radiative coupling
between neighbouring cells i.e. the fraction of emitted photons that is
absorbed in each of the neighbouring cells. Dust temperatures are solved from
a non-linear set of equations where the $i$:th equation contains three unknown
temperatures, one for the $i$:th cell and one for each of its neighbours. We
have used a simple iteration where one temperature at a time is updated until
all temperatures have converged. The calculation converges with very few
iterations, and time spent in solving these equations is insignificant
compared with the overall run-times. The use of tridiagonal operators is not
necessarily restricted to one-dimensional models. It could be used in
three-dimensional models where, due to the geometry, each cell interacts
mainly with two neighbours (e.g., pieces of spherical shells). 

The implementation of the AMC methods is described in more detail in 
Appendix~\ref{sect:app_amc}.

\section{Tests with one-dimensional models} \label{sect:1d}

In this section we test in practise the methods listed in the previous
section. Improvements in the sampling of the radiation field and in the
convergence of dust temperature calculations are studied separately. Tests are
made using spherically symmetric model clouds and assuming dust at an
equilibrium temperature.

The first model is taken from Stamatellos et al. (\cite{stamatellos}; their
model BE2). The cloud represents a sub-critical, externally heated spherical
cloud. The radial density distribution follows the Bonnor-Ebert solution for an
isothermal spherical cloud bounded by external pressure. In connection with
this cloud we use the dust model of Ossenkopf \& Henning (\cite{oh94};
coagulated grains with thin ice mantles accreted in 10$^5$ years at a density
of $10^6$cm$^{-3}$) and the external radiation field as given by Black
(\cite{black94}). Therefore, our calculations correspond exactly to those
presented by Stamatellos et al. The visual optical depth to the centre of the
cloud is 16.6. In the following this is called the model $S_1$. We study also
an optically thicker cloud, model $S_2$, where the optical depth has simply
been scaled up by a factor of ten.  Model clouds are divided into 50 shells of
equal geometric thickness and the number of simulated frequencies was 44.

From Niccolini et al. (\cite{niccolini}) we take a model where the optical
depth to cloud centre is 10 at 1\,$\mu$m. The cloud has a central radiation
source with a spectrum corresponding to a black body with $T$=2500\,K. We
study also a second model, $N_2$, which is identical to $N_1$ but has an
optical depth $\tau$(1\,$\mu$m)=100. The models consist of 51 shells. The
radiae are equidistant on logarithmic scale except for some very narrow shells
close to the cavity surrounding the central source. In accordance with
Niccolini et al. (\cite{niccolini}) we use a simple dust model where
absorption and scattering cross sections are proportional to the frequency and
scattering is isotropic. The number of simulated frequencies was 40.

\subsection{Sampling in temperature calculations} \label{sect:sampling_T}

\begin{figure}
\resizebox{\hsize}{!}{\includegraphics{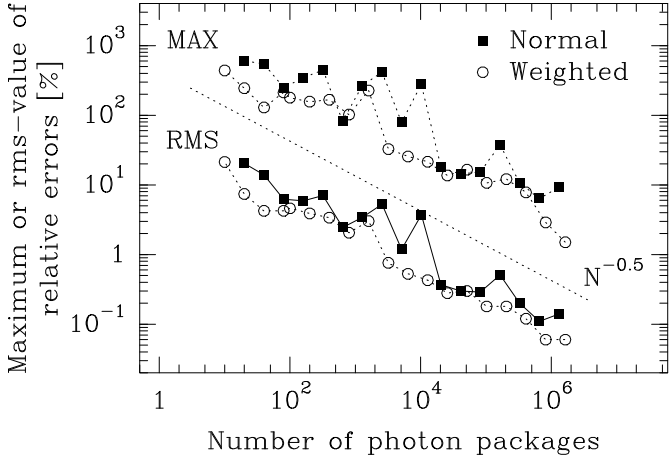}}
\caption[]{Convergence of calculated dust temperatures in model $S_1$ as the
function of the number of photon packages per iteration and frequency. Results are shown for
both normal Monte Carlo calculations (filled squares) and calculations where
weighting was applied to both dust emission and background emission (open
circles; see text for details). The upper curves show the maximum relative
error in any of the shells and the lower curves show the rms-value of the relative
errors summed over all shells.
The dotted line shows the expected $N^{-0.5}$ convergence rate.}
\label{fig:stam_sampling}
\end{figure}

Fig.~\ref{fig:stam_sampling} shows convergence as the function of the number
of simulated photon packages per iteration (and frequency), $n_{\rm p}$, for
model $S_1$. 
Maximum error and overall rms error are shown for derived dust temperatures in
different shells. The errors were determined by comparison with a calculation
with much higher $n_{\rm p}$. Results are plotted for normal Monte Carlo
sampling (no weighting) and for calculations where equal numbers of external
photon packages were sent towards each annulus described by the radial
discretization (weighting scheme $A$) and equal number of photons packages
were sent from within each cell (scheme $B$). Equal number of photon packages
were used for describing the external field and emission within the cloud. The
convergence is for both methods $\sim 1/\sqrt{n_{\rm p}}$ and differences in
accuracy remain relatively small. However, for weighted sampling the errors
are always equal or smaller than in normal runs. In particular, fluctuations
of the results are smaller and for some $n_{\rm p}$ the rms error is almost
one order of magnitude smaller than in runs where weighting was not applied. 

\begin{figure}
\resizebox{\hsize}{!}{\includegraphics{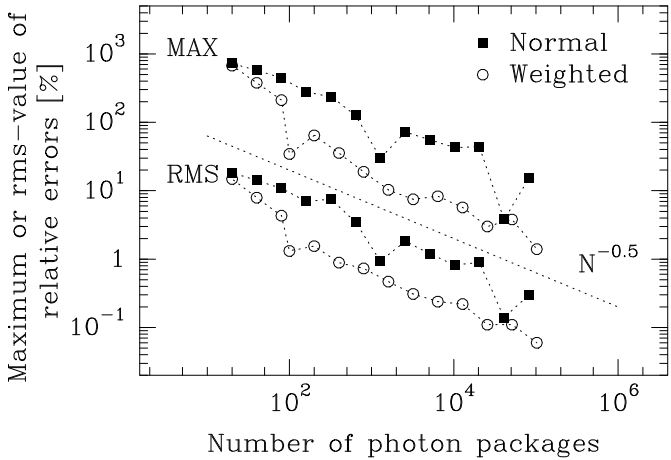}}
\caption[]{Convergence of calculated dust temperatures in model $S_2$ vs.
number of photon packages per iteration and frequency. The symbols are as in
Fig.~\ref{fig:stam_sampling} } 
\label{fig:stam_T_sampling}
\end{figure}

Fig.~\ref{fig:stam_T_sampling} shows same relations for model $S_2$ that has 
a factor of ten higher optical depth. In this case the advantage of weighted
sampling is more noticeable. On the average, errors have decreased `only' by a
factor of $\sim$8 but this translates to an  almost two orders of
magnitude difference in run times.

Models $N_1$ and $N_2$ are fundamentally different because of the internal
source. Both models are optically thick and dust temperatures are mostly
determined by dust emission itself. For optically thin models the default 
scheme could be improved only by making sure that photon packages are sent as
uniformly as possible towards different directions. This could be accomplished
with the use of quasi random numbers or equally by using pre-selected,
angularly equidistant directions.

\begin{figure}
\resizebox{\hsize}{!}{\includegraphics{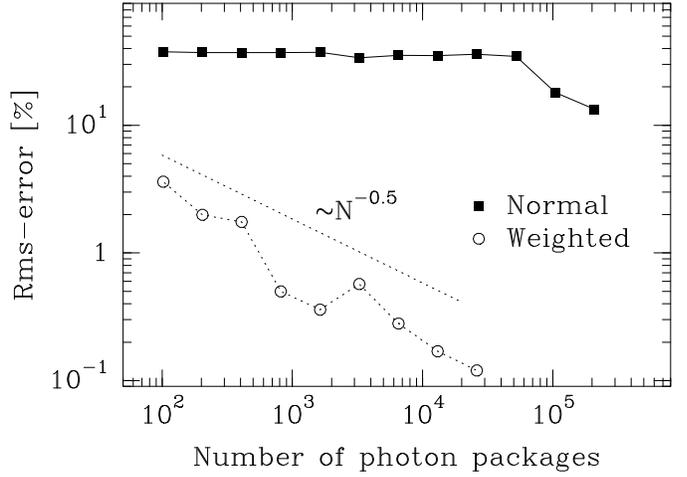}}
\caption[]{
Rms-value of relative dust temperature errors as function of the number of
simulated photon packages per iteration and frequency. Results are shown for
model $N_1$ for the normal Monte Carlo simulations (filled squares) and for
weighting scheme $B$ where relatively more photon packages are created close
to the centre of the model (open circles; see text for details). 
} 
\label{fig:tau10_sampling}
\end{figure}

The optical depth towards the centre of cloud $N_1$ is 10 at 1$\mu$ and the
visual extinction is roughly twice this value. The angular distribution of
photons emitted from the source is not weighted, background photons are not
included at all and we test only the effect of weighting the distribution of
locations at which dust emitted photon packages are created within the cloud
(method $B$). Results are shown in Fig.~\ref{fig:tau10_sampling} where
relative rms-errors from un-weighted and weighted runs are compared. If 
photon packages are created uniformly over the cloud volume the sampling of the
innermost shells remains very poor. The results are essentially incorrect
unless the number of photon packages is larger than the ratio between the
volume of the cloud and the volume of the smallest cells or rather the volume
of smallest optically thick region that could separate the outer cloud from
the source. The observed convergence of un-weighted results indicates that the
thickness of this inner region is in this case a few per cent so that only
about one photon package out of 10$^5$ samples emission from there. Results
show clearly the need of some kind of weighted sampling.

\begin{figure}
\resizebox{\hsize}{!}
{\includegraphics{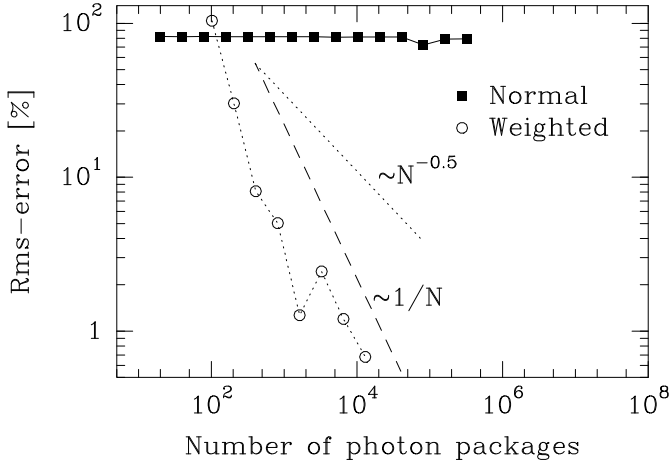}} 
\caption[]{
Convergence of dust temperature calculations in model $N_2$ as the function of
the number of photon packages per iteration and frequency (symbols as in
Fig.~\ref{fig:tau10_sampling}) 
}
\label{fig:tau100_sampling}
\end{figure}

In model $N_2$ optical depths are higher by a factor of ten and normal Monte
Carlo sampling becomes even more inefficient. This is clearly seen in
Fig.~\ref{fig:tau100_sampling}) where no convergence is seen before the number
of photon packages is $\sim$10$^6$ and adequate sampling of the inner region
becomes possible. For the weighted sampling the situation is much better and
actually not worse than in the case of the previous model $N_1$. As a result,
a given accuracy is achieved with a number of photon packages that is a factor
of 10$^4$ lower than in the case of unweighted sampling. Another remarkable
fact is that the convergence seems to be much faster than the usual
$1/\sqrt{N}$ behaviour. This is not altogether surprising since for regular
sampling (as accomplished in Monte Carlo integration by the use of
quasi-random numbers; see eg. Press \& Teukolsky \cite{press_nr},
\cite{press89}) the convergence is expected to be $\sim1/N$ rather than
$\sim1/\sqrt{N}$. In our case the regularity is restricted to systematic
selection of shells and while positions and directions within a shell are
completely random this is enough to ensure the faster convergence rate.

Previously, we used two weighting schemes ($A$ and $B$) in comparison with the
default Monte Carlo sampling.  As discussed in Sect.~\ref{sect:sampling} there
are at least two further ways to influence the sampling, especially in the
cloud centre where there would otherwise be very few photon packages. One can
send photon packages from cells preferentially towards the cloud centre
(scheme $C$), and scattered photon packages can be directed preferentially in
that direction (scheme $D$).
We tested the effect of these methods for models $S_1$, $S_2$, $N_1$, and
$N_2$, with the number of photons packages (per iteration and frequency) equal
to $n_{\rm p}=2000$. The rms values of the relative temperature errors over
all cells are shown in Table~\ref{tab:weighting} for some combinations of the
four weighting schemes.

\begin{table}
\caption[]{
Errors of computed temperature for models $S_1$, $S_2$, $N_1$, and $N_2$ when
default Monte Carlo sampling or different weighting schemes ($A$, $B$, $C$,
and $D$) are used. The numbers are rms values of relative errors given as
percentages. Number of photon packages per iteration was 2000. 
}
\begin{tabular}{lllll}
\hline
weighting scheme &    $S_1$    &    $S_2$    &    $N_1$    &    $N_2$     \\
none      &    3.2      &    1.7      &     34      &    81        \\
$A$       &    1.1      &    0.34     &     -       &    -         \\
$B$       &    3.2      &    1.7      &     0.49    &    4.7       \\
$C$       &    3.4      &    1.4      &     36      &    81        \\
$D$       &    1.7      &    1.7      &     37      &    82        \\
$A,B$     &    1.1      &    0.46     &     -       &    -         \\
$A,B,C$   &    1.1      &    0.31     &     -       &    -         \\
$A,B,D$   &    0.46     &    0.48     &     -       &    -         \\
$A,B,C,D$ &    0.46     &    0.31     &     -       &    -         \\
\hline
\end{tabular}  \label{tab:weighting}
\end{table}

For the externally heated models $S_1$ and $S_2$ the scheme $A$ (weighted
generation of background photon packages) was the most useful one and  it
reduced rms errors by a factor of $\sim$3 in model $S_1$ and by a factor of
$\sim$5 in model $S_2$. Methods $B$ and $C$ (distribution of positions and
directions of photon packages created inside the cloud) affected the results
only little. This is not surprising since in these models the dust emission has but
a small contribution to the radiation field.  Method $C$ produced small
improvements only in cloud $S_2$ while method $D$ was effective only in model
$S_1$. The best combinations were $A$, $B$, and $D$ in model $S_1$ and $A$,
$B$, and $C$ in model $S_2$, and the rms errors were correspondingly reduced by
factors $\sim 7$ and $\sim 6$.

In internally heated clouds $N_1$ and $N_2$ the emission from innermost small and
hot cells must be properly sampled - otherwise the flow of the re-emitted energy
through the optically thick cloud is cut and results will be incorrect. 
Table~\ref{tab:weighting} shows that errors are very large if scheme $B$ is not
used.  Scheme $A$ was not applied since background photons were not simulated.
The other weighting methods have no real effect on the errors. Methods $C$ and
$D$ aim at improving in cloud centre the sampling of radiation that originates
further out. In clouds $N_1$ and $N_2$ the flux of energy is mostly in the
reversed direction and the use of these methods is not helpful.

\subsection{Sampling of scattered photons}

The weighting of the angular distribution of scattered photon packages (scheme
$D$) can be used to improve sampling in selected regions inside the cloud. It
can also be used to improve the statistics of scattered flux observed outside
the cloud. For three-dimensional models the out-coming flux might be needed
only for one direction. In that case, scheme $D$ can be used to drive photon
packages towards the selected direction. The statistical noise will be
decreased for the scattered flux and the same simulation can still be used
for solving the temperature structure of the cloud.

We calculated for model $S_2$ 400 pixel maps of scattered flux at 0.55$\mu$m.
The method of forced first scattering was used in all calculations. In the
weighting scheme $D$ the probability function of the angular distribution of
scattered photons was $p\sim exp(-\gamma \theta)$ where the angle $\theta$ is
measured from a vector pointing towards the observer.  A reference solution
was obtained with normal Monte Carlo using 4$\times 10^7$ photons. In this map
the expected error per pixel is $\sim$2\%. Runs with 2$\times 10^5$ photon
packages were compared with this. The normal Monte Carlo method resulted in a
standard deviation of 34\%.  The corresponding error when using scheme $D$ was
$\sim 20$\% when the parameter $\gamma$ was in the range 0.3-1.0. For a given
noise requirement and $1/\sqrt{N}$ noise dependence the difference corresponds
to almost a factor of three in run times. One can expect that this kind of
weighting could be more useful in the case of non-isotropic scattering or
a non-isotropic radiation field.



\subsection{Convergence tests}

The methods of Sect~\ref{sect:accel} and their effect on the convergence of
dust temperature calculations were tested first with model $N_2$.  In that
model the optical depth to the cloud centre is $\tau$(1$\,\mu$m)=100.
Calculations were started with a flat temperature profile, $T=17$\,K, while in
the final solution the temperatures range from $\sim$140\,K to over 2000\,K.
Random number generators were reset after each iteration and the results were
compared with similar calculations with a very large number of iterations.
Therefore, the results do not show any sampling errors and reflect only the true
convergence of dust temperature values.

\begin{figure}
\resizebox{\hsize}{!}{\includegraphics{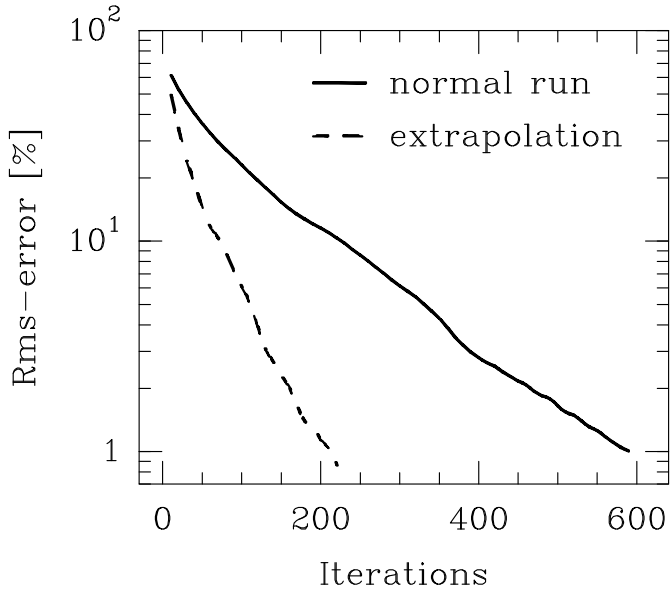}}
\caption[]{
Convergence of dust temperatures in model $N_2$ in the case of normal
iterations and with the use of the extrapolation method. The plotted values
are rms-values of the relative temperature errors.
}
\label{fig:ng}
\end{figure}

Fig.~\ref{fig:ng} shows the rms-errors as function of the number of iterations
for two runs. With the normal method the convergence is quite slow and a
relative rms error of 1\% is reached only after $\sim$600 iterations. When the
extrapolation method of Sect.~\ref{sect:accelxpol} is used the convergence
improves by a factor of three and an accuracy of 1\% is reached after some
200 iterations. The extrapolation step was done always after three normal
iterations. The convergence is quite smooth and extrapolations did not at any
point produce noticeable oscillations. This suggests that the convergence rate
could be still improved by amplifying the computed corrections, at least
during the first iterations.

\begin{figure}
\resizebox{\hsize}{!}{\includegraphics{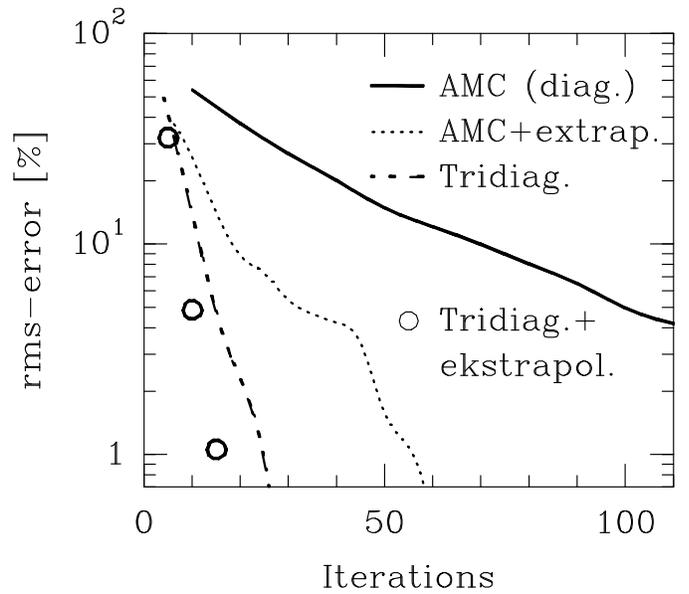}}
\caption[]{
Convergence of dust temperatures in model $N_2$ for diagonal AMC-method (solid
line), diagonal AMC with extrapolation (dotted line) and AMC with tridiagonal
operator (dash-dotted line).
}
\label{fig:ali}
\end{figure}

Results from corresponding tests with the AMC-methods are shown in Fig.~\ref{fig:ali}.
With the diagonal operator (i.e. when internal absorptions are treated
explicitly) the convergence rate is about the same as with the extrapolation
method and a relative accuracy of 1\% is reached with $\sim$200 iterations.
When the two methods are applied simultaneously this accuracy is reached with
60 iterations. The convergence rate is very good both initially and again after
some 45 iterations. In between the rate is slower, possibly because of some
less successful extrapolation steps. In this test the tridiagonal operator is
clearly better with a good and constant convergence rate. Finally, the circles
in Fig.~\ref{fig:ali} show the rms error after every fifth iteration when, in
addition to the tridiagonal operator, an extrapolation step is done after every
fifth iteration. This gives the fastest convergence and an rms error of 1\% is
reached with just 15 iterations, a factor of 40 less than for the basic
method.

\subsection{Reference field}

\begin{figure}
\resizebox{\hsize}{!}{\includegraphics{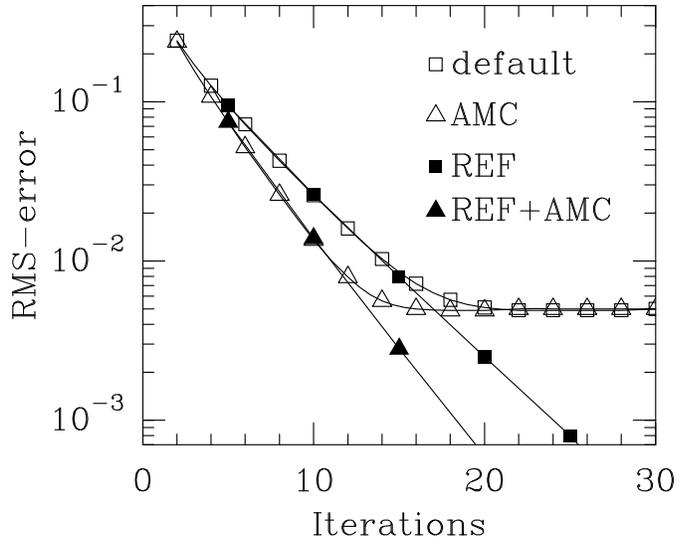}}
\caption[]{
Convergence of dust temperatures in model $N_1$ with (filled symbols) and
without (open symbols) the use of a reference field when the number of photon
packages per iteration is equal. The rms-values of relative temperature
errors are plotted as function of the number of iterations.
}
\label{fig:reftau10}
\end{figure}

The reference field concept was tested with model $N_1$ which is more
optically thin than $N_2$ but where the solution still requires many
iterations. Figure~\ref{fig:reftau10} shows the results obtained with and
without accelerated Monte Carlo and with and without a reference field. In all
calculations the number of photon packages per iteration and frequency was
4080, half of which described flux from the central source while the other
half described the dust emission. The reference solution corresponded this
time to calculations with both a larger number of iterations and photon packages
per iteration. Because of the smaller optical depth the advantage of
accelerated Monte Carlo is smaller than in model $N_2$ (see Fig.~\ref{fig:ali}
above) but still clear. If the reference field is not used the final accuracy
depends on sampling errors i.e. on the number of photon packages per
iteration. When a reference field was not employed the same set of
random numbers was used on each iteration. This explains the smoothness of the
horizontal curve. If different random numbers were used the error level would
vary from iteration to iteration but the general level would remain the same.

In principle, the use of a reference field does not affect the rate of
convergence. However, after the first iteration no emission from the central
source is simulated since that information is already contained in the
reference field. The following iterations should be about twice
as fast if the overhead from the use of a reference field is ignored. In fact,
measured after ten iterations the ratio of run times is 2.9:1 in favour of the
reference field. Photon packages emitted from the central source are inside
the most optically thick region and therefore take a longer time to simulate.
This explains why the ratio is in this case even larger than 2:1. The second
advantage of a reference field is that the sampling improves from iteration to
iteration and the final errors depend on the total number photon packages
simulated. In the case of Fig.~\ref{fig:reftau10} this means that convergence
remains linear below the level that was reached without the use of a reference
field. Emission from the central source was in this case simulated with 2040
photon packages per frequency while for dust emission the number of packages
was 2040 times the number of iterations. It is clear that in calculations
without a reference field the accuracy was limited by the sampling of the dust
emission.

When the reference field is used the improvement as a function of the number
of iterations depends both on convergence of temperatures and improvement of
sampling. In the test shown in Fig.~\ref{fig:reftau10} the number of packages
per iteration was clearly sufficient so that the improvement of accuracy did
not at any point decelerate because of sampling errors. Similarly, 2040 photon
packages from the central source at each frequency were sufficient so that
those sampling errors are not yet visible in Fig.~\ref{fig:reftau10}.  One
could easily use more photon packages to simulate emission from sources and
background since that simulation is done only once. If the number of photon
packages describing dust emission is too low it slows down the convergence but
does not necessarily prevent it. In order to minimise the run times one should
use as small a number of photon packages as possible. In the case of model
$N_1$ the number could be brought down to $\sim$100 without significantly
affecting the convergence rate seen in Fig.~\ref{fig:reftau10}.

\section{Discussion}  \label{sect:discussion}

\subsection{Monte Carlo sampling}

Several weighting schemes were used to improve the sampling so that a given
accuracy could be reached with fewer photon packages and shorter run times.
All four schemes $A$, $B$, $C$, and $D$ were useful in some conditions. For
optically thin or moderately thick and externally heated clouds the scheme $A$
(weighting of spatial distribution of background photons) was the most useful
one while for optically thicker and internally heated clouds the scheme $B$
was by far the most important one. The overall reduction in the number of
photon packages ranged from a factor of a few to over $\sim$10$^4$. The
results are extremely model dependent. However, a brief analysis of the model
structure and the expected flow of energy within the model will give a good
idea on what kind of a weighting scheme would be most useful.

Stamatellos \& Whitworth (\cite{stamatellos}) sent external photons into the
model cloud from one location at the cloud edge with probability distribution 
$p(\theta)\sim sin\theta cos\theta$.  This distribution does not give a
particularly good sampling towards the cloud centre ($\theta\approx \pi$). 
The weighting scheme $A$ ensured that the effect of the external field could
be calculated accurately in the cloud centre no matter how the radial
discretization was done. This would become more important if the number of
shells were increased or the size of the innermost shells otherwise decreased.
The systematic sampling used in the method (equal number of packages towards
each annulus) also decreases Monte Carlo noise. As a consequence, the run times
become rather short. For $S_1$ a relative 1\% accuracy of temperatures was
reached with 3000 photons per frequency. Half of the photon packages was sent
from the background and half from cells within the model cloud. We did not
try to optimise this ratio. With an 600MHz PIII computer the run time was
$\sim$8 seconds per iteration.

The method $A$ should be most important for optically thin clouds but it was
also found useful for the optically thick cloud $S_2$. One explanation is that
while short wavelength radiation is absorbed in the outer layers the cloud
centre is heated by longer wavelength radiation for which the optical depth
may be close to unity. The spatial distribution created for external photons
at the surface of the cloud is not completely destroyed by scatterings and
will still result in more accurate estimates of the heating in inner regions.
Details depend, of course, on the dust model and the spectrum of the external
radiation field. 

For internally heated models $N_1$ and $N_2$ the method $B$ was indispensable.
In the case of model $N_1$ an accuracy of 1\% was reached with just a few
hundred photon packages per frequency. With $10^3$ packages per iteration and
frequency the run time with a 600MHz PIII computer was down to $\sim$8 seconds
per iteration. The required number of iterations can be read from
Fig.~\ref{fig:reftau10}. If the reference field is used the run time drops by
more than half and can be shortened further by decreasing the number of
simulated photon packages.  We repeated the calculations with accelerated
Monte Carlo (diagonal operator) and using a reference field, with 100 photon
packages per frequency per iteration for the dust emission and 1000 photon
packages for the central source in the first iteration. An rms-error of 1\%
was reached after 10 iterations and the whole calculation took less than 12
seconds. If these run times are compared with Niccolini et al. \cite{niccolini}
one should also note that while in our runs the cloud was divided into 51
shells they used 65 shells (and 20 additional shells in the empty inner cavity).

Method $C$ (weighting of angular distribution of dust emission) did produce
small improvements only in model $S_2$ but method $D$ was not particularly 
useful in any of the four models. This is partially due to the selection of
the test cases. For example, in models $N_1$ and $N_2$ the source is situated
in the cloud centre and the flow of energy is outwards while method $D$ tried
to improve the sampling of the scattered flux flowing in the opposite
direction. Method $D$ could still be useful, e.g., in cases where a cloud is
heated from outside by a nearby star. 

A change in the sampling can affect the average time it takes to simulate a
photon package. In an optically thin and spherically symmetric model those
photons that are sent towards cloud centre pass through most cells and take
more time to simulate. Frequent scatterings make simulations more time
consuming in optically thick parts of the clouds. Therefore, weighting schemes
of Sect.~\ref{sect:sampling} should be compared not only against the number of
photon packages required but also against the actual run times. Some tests
were made with the optically thick model $N_1$. With weighting schemes $A$ and
$B$ the average run times per package were only $\sim$3\% longer than in the
case of unweighted Monte Carlo. The increase is insignificant compared with
the reduction in the required number of photon packages. For model $S1$ the
difference in run times between normal Monte Carlo and weighting scheme $A$
were similarly about 3\%.

\begin{figure}
\resizebox{\hsize}{!}{\includegraphics{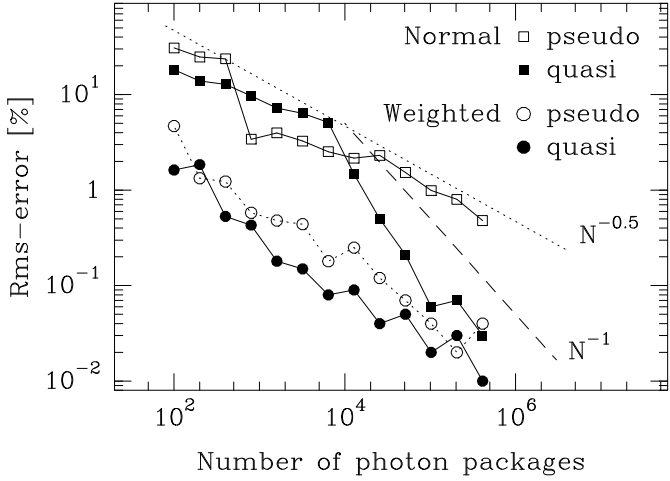}}
\caption[]{
Convergence as function of the number of photon packages for model $S_1$ with
normal (squares) and weighted Monte Carlo (circles). Results obtained with
pseudo random numbers (open symbols) and quasi random numbers (filled symbols)
are shown
}
\label{fig:stam_sampling_forward}
\end{figure}

The use of quasi random numbers was shortly tested but no significant
improvement was observed when compared with results obtained with pseudo
random numbers.  For the unweighted case the results from separate runs became
more consistent (i.e. fluctuations decreased as seen in, e.g.,
Fig.~\ref{fig:stam_sampling}). The average error level was not improved and
convergence rate (as a function of the number of photon packages) remained
proportional to $1/\sqrt{n_{\rm p}}$. The use of quasi random numbers is
straightforward for photon packages coming from the background and from
internal sources. For emission within the cloud the position and direction of
the created photon packages must be coordinated i.e. these should be created
from a quasi random number sequence in five dimensions (three coordinates for
position and two for direction). In our tests a separate random number
sequence was used for each emission source and for the scattering process.
However, it is still possible that scatterings destroy the equidistant
sampling and could explain the slow convergence. 
This was checked by repeating the calculations for model $S_1$ assuming pure
forward scattering.  Absorptions are calculated explicitly along the photon
path so that results are not affected by the positions of the scatterings.
Fig.~\ref{fig:stam_sampling_forward} shows convergence for un-weighted and
weighted sampling (methods $A$ and $B$ applied). This shows that in normal
Monte Carlo calculations the convergence rate does finally become $\sim 1/N$
and after some $10^5$ photon packages the accuracy reaches that of the
weighted calculations. When weighted sampling is used quasi random numbers
give a slightly lower error level but the convergence remained at $\sim
1/\sqrt{N}$.
It may still be possible to improve the uniform sampling of scattered flux. This could
be accomplished by using separate random number sequences in different parts
of the cloud but there may be other, more practical methods.  The possibility
of ultimately reaching convergence proportional to $1/N$ makes further studies
necessary.

\subsection{Convergence}

Tests showed that simple extrapolation based on previous temperature values was
able to decrease the overall run times by a factor of a few. The method requires
that temperature values are stored for two previous iterations. This extra
memory is, however, inconsequential for one-dimensional models and is
even in three-dimensional models only a small fraction of the total memory
requirements.

For optically thick clouds and when self-coupling between dust temperatures in
different parts of the cloud is important AMC-type methods provided equally
good or better results. The use of the tridiagonal operator is restricted
mostly to one-dimensional models, but diagonal operator can be applied equally
well in three-dimensional models. The use of the diagonal operator requires
storage of photon escape probabilities, i.e. one number per cell and frequency.
This doubles the amount of data produced during the simulation of the
radiation field. If one takes into account all data that are needed in the
calculations (optical depths for absorption and scattering, source functions
etc.) the percentual increase is not very significant. For tridiagonal operator two
additional numbers per cell and frequency need to be stored since calculations
require information on the coupling between neighbouring cells. This could
possibly be managed even in three-dimensional case. However, there the method
is useful only if each cell interacts mainly with two other cells. This might
be the case, e.g., for cylindrically symmetric models with thin cylinders or in the
case of a three-dimensional model built of pieces of spherical shells.
Depending on the geometry (and the available memory) it may be useful to
include interactions between several neighbouring cells. The implementation is
not essentially more complicated than for the tridiagonal operator.

The extrapolation method and AMC-methods introduce some overhead the
importance of which depends on the number of simulated photons packages, cells
and frequencies. For the extrapolation method the simulation part is identical
with the default method. The time required for the extrapolation is small and
since extrapolation is done, for example, only every third iteration the
increase in run times per iteration is unnoticeable. For AMC-calculations with
diagonal operator the computations differ again very little and no differences
in run times per iteration was observed. For the tridiagonal operator the
situation is somewhat different. In our implementation dust temperatures were
solved in each iteration from a set of nonlinear equations and this solution
was also an iterative one. In spite of this it was difficult to see any
difference in run times per iteration. For example, with model $N_2$ and even
in the first iterations when the solution was far from the correct one, and
more sub-iterations were presumably needed to solve the set of equations, the
difference in the total run time per iteration was at most $\sim$1\%.

\subsection{Comparison with earlier studies} \label{sect:comparison}

In the case of the model $S_1$ from Stamatellos \& Whitworth
\cite{stamatellos} and the model $N_1$ from Niccolini et al.\cite{niccolini}
our results are in perfect agreement with the results given in those articles.
Niccolini et al. presented timings for their calculations of the model $N_1$.
They found an rms-error below 1\% when number of photon packages was a few
times $10^5$ while in our Fig.~\ref{fig:tau10_sampling} the rms-error for the
{\em un-weighted} Monte Carlo method was still at least above 10\%. The
difference is probably caused by a difference in the radial discretization.
Niccolini et al. \cite{niccolini} used 65 shells and although we have only 51
shells the innermost shells have relatively much smaller volumes. Therefore,
if normal Monte Carlo sampling is used the temperature errors remain very
large in these shells and the rms-errors of Fig.~\ref{fig:tau10_sampling}
reflect this fact. When weighted Monte Carlo was used the effect of the
different volume sizes is removed and 1\% accuracy was reached with $\sim10^4$
photon packages per iteration. Comparison of run times is even more uncertain
because of the different platforms used.
For their method 2 and for $2.4\times 10^6$ photons Niccolini et al. reported
a run time of 500\,s on a Cray T3E-1200 parallel computer with 16 processors.
Since an accuracy of 1\% was reached with a few times $10^5$ photons those
computations would have taken $\sim 50$s. With accelerated Monte Carlo and
weighted sampling our calculations took a total of 12 seconds on a 600MHz PIII
computer. It is not clear whether the run times given by Niccolini et al. were
for one iteration or the whole run. Nevertheless, the comparison seems to
indicate that the use of weighted sampling and AMC-type acceleration can
indeed result in significant savings.

%
The method of Bjorkman \& Wood (\cite{bjorkman01}; in the following BW) was
described to solve the dust temperatures without iterations and therefore the
equilibrium temperature calculations would require no more time than a pure
scattering model.  The BW method does, indeed, have several advantages over
the default Monte Carlo method. Firstly, all simulated photon packages are
used in the derivation of the final solution. In normal Monte Carlo runs the
radiation field is estimated on each iteration independently using a new set
of photon packages and, compared with the total number of simulated packages,
the random errors are larger. Secondly, in the BW scheme corrections are
applied immediately when a new photon is absorbed. In normal Monte Carlo,
cells continue to send photons corresponding to the old temperature until all
temperatures are updated at the end of the iteration. Finally, in the BW
scheme the photons follow the actual flow of energy so that more emissions
(and re-emissions) take place in the warmer region. Therefore, the sampling of
these regions is particularly good. The immediate re-emission mechanism helps
to ensure that there will be no poorly sampled regions that could separate
energy sources from the rest of the model (see Sect.~\ref{sect:sampling_T}).
When a short-wavelength photon is absorbed (e.g., close to a radiation source)
this is usually followed by a re-emission at a longer wavelength for which the
optical depths are likely to be considerably lower. The resulting increase in
the photon free path makes the simulation procedure much more efficient in the
case of large optical depths. However, most of the other listed deficiencies
of the normal Monte Carlo method can be alleviated by using weighted sampling
and a reference field.

In the BW method energy conservation is enforced for each photon package
separately. Once the simulation of the selected number of photons from the
various radiation sources has been completed the dust temperatures will also
have reached -- apart from sampling errors -- their final, correct values. It
was argued that in the case of high optical depths the method does not suffer
from similar slowdown as $\Lambda$-iteration methods. This does not mean that
run times would not be affected by $\tau$. Consider, for example, a cell which
is optically thick even for the re-emitted radiation. Successive re-emissions
and absorptions slowly increase the temperature of this one cell while the
flow of information across cell boundaries is minimal. Each absorption implies
a new calculation of the dust temperature so that, like in $\Lambda$-iteration
methods, new dust temperatures must be solved numerous times. The efficiency
of these temperature updates is clearly crucial for the overall run times. The
BW implementation used large arrays of pre-tabulated Planck mean opacities to
speed up the solution (see Bjorkman \& Wood \cite{bjorkman01}, Eq. 6). 

In the default Monte Carlo scheme information is first gathered from a large
number of photon packages and new dust temperatures are calculated only at the
end of an iteration. Therefore, the ratio between the number of simulated
photon packages and the number of temperature updates is much larger than
in the BW method. This will become significant if calculations include
several dust populations, grains size distributions or transiently heated
grains. Against this background, it is interesting to see a comparison of the
methods in the simplest case of one dust population, one grain size and grains
at one equilibrium temperature. We compared our run times with the program
MC3D\footnote{The program MC3D can be downloaded at
http://www.mpia-hd.mpg.de/FRINGE/SOFTWARE/mc3d/} by S. Wolf (\cite{wolf03})
which includes an implementation of the BW method. The cloud model consists of
central black body source with radius $R_{\rm S}$ and temperature 2500\,K.
Beyond a central cavity the surrounding cloud extended from $3\,R_{\rm S}$ to
$300\,R_{\rm S}$ and had a density distribution $n\sim r^{-2}$.  The dust
consists of 0.12\,$\mu$m astronomical silicate grains (Draine \& Lee
\cite{dl84}). We consider two models where optical depth to cloud centre is
either 100 or 1000.

\begin{figure}
\resizebox{\hsize}{!}{\includegraphics{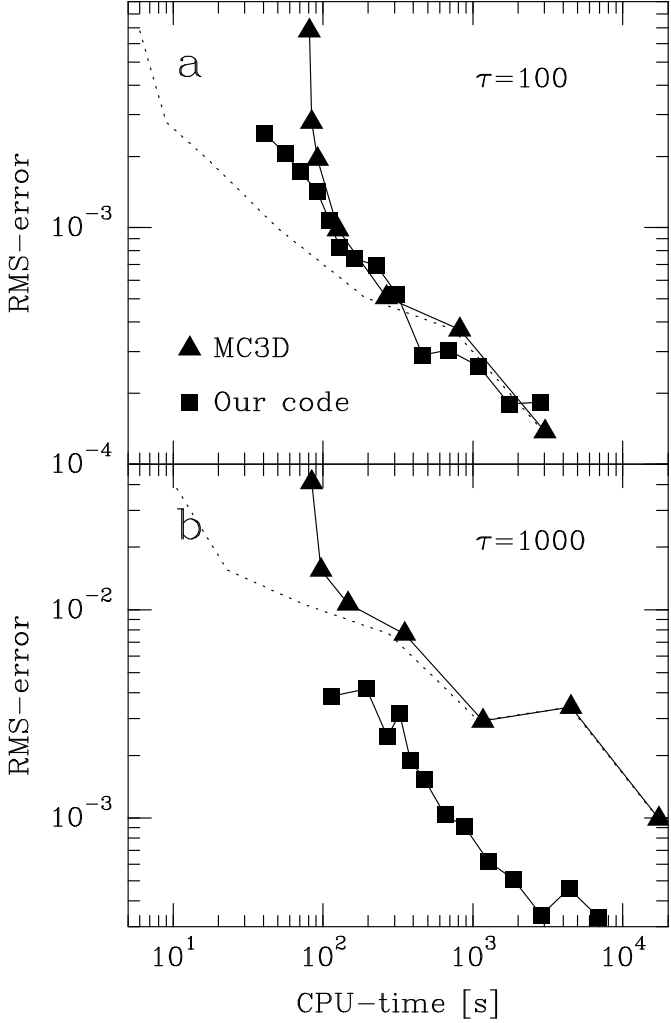}}
\caption[]{
Comparison of run times between our program (solid squares) and the Bjorkman
\& Wood (\cite{bjorkman01}) method (triangles) as implemented in the MC3D
program of Wolf (\cite{wolf03}). The figures show rms-errors of the computed
dust temperatures as functions of CPU-time. The two frames correspond to
models with central optical depth $\tau_{\rm V}=100$ and $\tau_{\rm V}=1000$.
The dotted lines indicate run times of MC3D program when the longer
initialisation time (pre-calculation of Planck mean opacities etc.) has been
subtracted.
}
\label{fig:comparison}
\end{figure}

In runs with the MC3D program only the number of simulated photons was changed
and otherwise default parameters were used (including code acceleration that
neglected very small fluxes). In our own program we used weighting scheme $B$,
a reference field (see Appendix~\ref{sect:app_reference}) and iterations were
accelerated with the AMC method using a diagonal operator. Both the number of
iterations and the number of photon packages per iteration were changed
between runs.  The rms errors were estimated by comparing obtained dust
temperatures with reference solutions that were computed separately for both
programs using a much larger number of photon packages (and iterations). The
rms-errors as the function of run times are shown in
Fig.~\ref{fig:comparison}. All runs were made on the same computer (AMD Athlon
MP 2000+) on a single processor.

%

%
Comparison of the methods themselves is affected by the usual uncertainties
(programs use different programming languages and the final run times depend
on details of the actual implementation and compiler optimizations). However,
some conclusions can be drawn. First of all, with the aid of a reference field
and AMC acceleration the normal Monte Carlo method is competitive with the BW
method. Secondly, the AMC method seems to have an advantage at higher optical
depths, in this case at $\tau\sim1000$. As noted above, the BW method requires
frequent temperature updates. If a normal Monte Carlo run is accelerated with
an AMC scheme the total number of temperature updates remains much lower. By
inference, in situations where temperature updates are more expensive the
combination of an AMC scheme and the use of a reference field may prove to be
more efficient than the BW method.

In addition to run times one may compare memory requirements of possible
implementations. If, as in the case of the previous example, dust properties
are identical in all cells the information of the general $\kappa_{\rm
sca}(\nu)$ and $\kappa_{\rm abs}(\nu)$ curves can be kept in memory. In the BW
scheme the local density and temperature would need to be stored for each cell
but the $\tau_{\rm abs}$ and $\tau_{\rm sca}$ values can be calculated based
on the local density and the interpolated or calculated $\kappa_{\rm
sca}(\nu)$ and $\kappa_{\rm abs}(\nu)$ values. Total memory requirement
corresponds to only $\sim2\times N_{\rm cells}$ floating point numbers. Let us
next considers a more complex problem with $N_{\rm pop}$ dust populations,
each discretized into $N_{size}$ size intervals. In this case memory is
required for $\sim N_{\rm cells} \times N_{\rm pop} \times N_{\rm sizes}$
temperature values. If dust abundances are not constant, there will be further
$N_{\rm cells} \times N_{\rm pop}$ abundance values.  Optical depths could be
calculated in real time summing over the dust populations although this could
have some negative impact on the run times. However, it seems probable that
during each temperature update one needs to have access to all absorption
cross sections so that the absorbed energy (calculated based on total optical
depths) can be divided between grains in different grain populations and size
intervals.  In order to relax the memory requirements one could read and write
temperature values directly to disk files but, since these values are updated
after each absorption, this would cause a significant increase in the run
times. Finally, if the problem includes transiently heated particles those
must be discretized into $N_{E}$ enthalpy bins. No implementation of the BW
scheme yet exists for non-equilibrium dust grains. However, in that case not
only are the temperature updates much more time consuming (relative to the
simulation part) but involve significantly larger amount of data, up to
$N_{\rm cells} \times N_{\rm pop} \times N_{\rm size} \times N_{E}$
temperature values.

In normal Monte Carlo the simulation of different discrete frequencies can be
made sequentially and completely independently. Each time the simulation of a
new frequency starts a table of the $\tau_{\rm abs}$ and $\tau_{\rm sca}$
values can be calculated for this one frequency and all cells. A priori, no
assumption need to be made that absorption and scattering cross sections would
be the same in all cells. Counters are also needed for photons emitted and
absorbed at the current frequency. Therefore, the total memory requirement is
$\sim 4\times N_{\rm cells}$ numbers. The AMC method with diagonal operator
adds to this $N_{\rm cells}$ numbers and, alternatively, a reference field
$\sim2\times N_{\rm cells}$ numbers. In runs with both AMC (diagonal operator)
and a reference field is $\sim7\times N_{\rm cells}$ floating point numbers
i.e. a few times the amount needed in the BW scheme. The program version used
in this paper follows this scheme apart from the fact that absorption an
emission counters are all kept permanently in main memory (this applied also
to the additional counters used with the reference field and the AMC methods).
This avoids the need to read and write tables of $n_{\rm abs}$ and $n_{\rm
emit}$ values between simulations of different frequencies. This does not
affect the run times of 1D models since all values fit in disk cache. We
tested the effect on run times in a more realistic setting, using a 3D model
where each cell was hit by $\sim100$ photon packages per iteration and
simulated frequency. The version where all $n_{\rm abs}$ and $n_{\rm emit}$
values were permanently in main memory was faster by $\sim$7\,\%. This shows
that the overhead associated with these disk operations is not very
significant. This is not surprising since each value is read and/or written
once per iteration but is consequently used in the simulation of possibly
hundreds of photon packages. Finally, let us consider a case where any of the
factors $N_{size}$, $N_{\rm pop}$, or $N_{E}$ are larger than one. During the
simulation one needs only total absorption and scattering cross sections for
one frequency at a time and the counters, $n_{\rm abs}$, $n_{\rm emit}$ etc.,
register the total number of events at that frequency. Therefore, memory
requirement of the simulation are not affected by any of the factors
$N_{size}$, $N_{\rm pop}$, or $N_{E}$. 

The traditional Monte Carlo scheme allows an efficient use of external files
and each value ($n_{\rm emit}$, $n_{\rm abs}$, etc.) is read from and written
to an external file exactly once per iteration.  The total overhead from all
file operations depends on the number of photon packages simulated but should
be clearly below 50\,\%. In the test mentioned above the overhead was 7\% per
file of $N_{\rm cell}$ elements. The reference field and AMC (diagonal
operator) require, in addition to $n_{\rm abs}$ and $n_{\rm emit}$ counters,
three such files. It might also be possible to calculate cumulative absorbed
energy without storing absorbed energy at each frequency separately. This
would further eliminate the need for storing $n_{\rm abs}$ values separately
for each frequency. In the BW scheme the random selection of the frequency of
the re-emitted photons means that optical depths must be calculated in real
time or large arrays are needed for pre-calculated values. Each absorption
event leads to a temperature update and this leads either to very frequent
disk operations (if temperatures are kept on disk) or large memory
requirements. In Fig.~\ref{fig:comparison} the comparison of run times was
based on one type of grains at one equilibrium temperature. In more complex
problems the ratio of run times will probably change (see above) and the ratio
between the memory requirements of a BW program and our program (AMC and
reference field included) will probably be of the order of $\sim (\times
N_{\rm pop} \times N_{\rm size} \times N_{E})/3$.

\subsection{Monte Carlo calculation at high optical depths}

Large optical depths cause two problems for radiative transfer calculations.
The sampling of the radiation field becomes more difficult as the photon free
path becomes very short and, secondly, iterations converge much more slowly.
The first problem can be addressed with weighted sampling and the second is at
least partially remedied with AMC methods. So far we have considered spherical
models with central optical depths up $\tau=1000$. The dust model is now
changed so that optical depth is $\tau_{\rm abs}=10000$ at {\em all} frequencies
and re-emission at longer wavelengths no longer provides an easier escape path
for the radiated energy. 

In an extremely optically thick cell most of the emitted energy is absorbed
within the same cell and photons packages only rarely move to another cell.
This means that one must create very large number of photon packages before
the energy flow between cells can be estimated with any accuracy. Weighted
sampling gives a solution for this particular problem. If photon packages are
created close to a cell boundary it is much more likely that it will
eventually cross it i.e. give information about the actual energy flow. For
example, one could use exponential probability distribution $q\sim
e^{-\tau(s)}$, where $\tau(s)$ is the optical depth to the closest cell
boundary. However, in this case the weights of individual photon packages are
random variables and the total emission from a cell would fluctuate. For
optimal results these fluctuations should be corrected, for example, by adding
a few photon packages for which the sum of photons corresponds to the
difference between the expected value and the number of photons so far simulated.

In the following we use a simpler approach. We assume that all photons further
than $\tau_{\rm abs}=15$ from the closest cell border are absorbed in the
cell. All simulated photon packages are started in the remaining volume where
distance to the closest border is below $\tau_{\rm abs}=15$. Weighting takes
into account the fact that packages are started only in a part of the cell
volume. Same number of photon packages in sent from each cell (weighting
scheme $B$) and the number of true photons simulated from each cell does not
fluctuate. The fraction $e^{-15}\approx 3\cdot 10^{-7}$ is sufficiently small
so that the true flux from deeper in the cell is insignificant compared with
the emission from regions closer to cell surface, $\tau(s)=0-15$. If optical
depth of a cell is increased beyond $\tau_{\rm abs}=30$ the accuracy of the
sampling will no longer be affected. Without AMC the convergence would be
extremely slow since the ratio between photons coming from neighbouring cells
is much smaller than the number of photons emitted and absorbed within a cell.
AMC scheme is used to eliminate the local emission-absorption cycle and
temperature updates depend only on photons that actually cross the cell
borders.

\begin{figure}
\resizebox{\hsize}{!}{\includegraphics{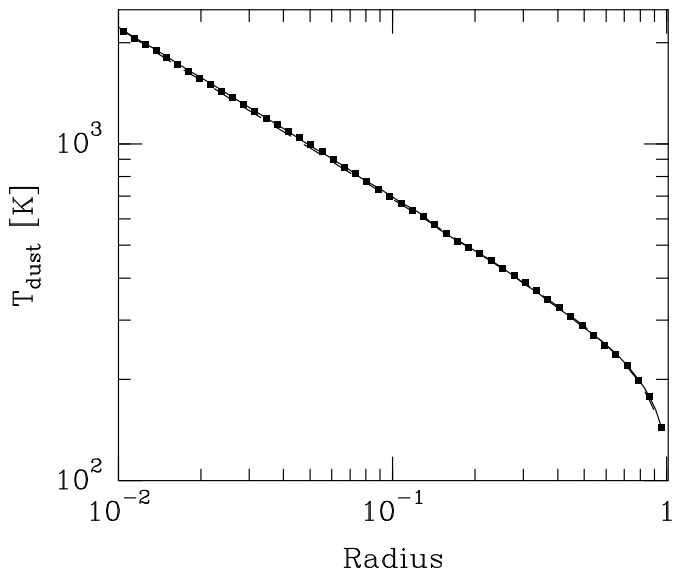}}
\caption[]{
Dust temperatures in a model with a central black body source and a
surrounding dust cloud with an optical depth $\tau=10000$ at all frequencies.
The solid line shows the results when emission was simulated only from those
regions where distance to the closest cell boundary was below $\tau_{\rm
abs}=15$ (see text). The solid squares correspond to a run with normal
weighting (scheme $B$) and a factor of 4 larger number of photon packages.
Direct numerical solution for a case where energy transfer across a surface is
$\propto A \times T_{{\rm dust}}^4$ (infinite optical depth) is drawn with a
dashed line.
}
\label{fig:tau_inf}
\end{figure}

Figure~\ref{fig:tau_inf} shows some results from our run (solid line). From
the dust medium 5000 photon packages were simulated each iteration. The
emission from the central star was divided between 1000 iterations using 500
photon packages per iteration. This was done to improve the use of the
reference field (see Appendix~\ref{sect:app_reference}). The total number of
iterations was 1500 and the run time $\sim$15 minutes. For reference we plot a
direct numerical solution for a case where energy transfer between cells would
depend only on cell area and temperature, $\propto A_i \times T_{{\rm
dust},i}^4$ (dashed line). This would correspond to infinite optical depth.
The comparison shows that a reasonable solution was indeed found in a
relatively short time. The small fluctuations in the computed dust
temperatures further indicate that the higher optical depths did not cause
numerical problems. In this case a similar solution could be reached even with
regular weighting (scheme $B$) using a minimum of $\sim20000$ photon packages
per iteration (solid squares). When sampling is restricted to regions close to
cell boundaries it is possible to increase the optical depths further. A
solution for $\tau=10^5$ model was found in same time and with the same number
of photon packages as for the $\tau=10^4$ model (5000 photon packages per
iteration and frequency for the dust emission). The resulting dust temperature
distribution was ($\sim$within noise) identical with the ones shown in
Fig.~\ref{fig:tau_inf}.

The previous example shows that the Monte Carlo method can, in principle, be
used even the case of cells with very large optical depths. Photon packages
are used only to sample the energy flux across cell boundaries and there are
not necessarily any packages that move across even a single cell. Information
moves across the cloud at the rate of one cell per iteration and run times
depend directly on the discretization of the optically thick region. The
weighted sampling must be tailored according to the cell geometry. In regular
geometries (e.g., a 3D cartesian grid) the implementation is relatively easy
and the weighting can always be restricted to regions where it is actually
needed. In optimal case the optically thick region would be treated separately
so that, in addition to weighting, even complete iterations could be carried
out independently from the rest of the model. Such methods are already in
preparation for hierarchical models. Hierarchical discretization makes it
possible to keep optical depths of individual cell lower and, if incoming
intensity is saved at grid boundaries, the calculations of the various
sub-grids can be carried out relatively independently. Consequently, a large
number of iterations required in certain sub-grids does no longer necessarily
have a large impact on the overall run times.


\section{Conclusions}

We have discussed various methods that can be used to improve the efficiency
of Monte Carlo radiative transfer calculations for dust scattering
and emission. The accuracy of the simulation of the radiation field can be
improved with weighted sampling and/or by employing a reference field. 
To our knowledge this is the first time that the reference field and the
weighted angular distribution of scattered and dust emitted photons are
considered in connection with dust continuum calculations. More surprisingly,
even spatial weighting of initial positions of photon packages is usually not
used.
Tests with internally and externally illuminated/heated spherical model clouds
lead to the following conclusions:

\begin{itemize}
\item improper sampling can lead to wrong results although calculations 
have some appearance of convergence 
\item 
weighted sampling resulted in significant savings in run times
ranging from a factor of $\sim1-10$ in externally heated models to several
orders of magnitude in internally heated and more optically
thick models
\item one should use weighting schemes that reduce fluctuations in the
sampling, e.g., simulating the same number of photon packages from each cell - this
reduces sampling errors and can lead to a faster convergence,
$\sim1/n_{\rm p}$, as function of the number of photon packages
\item 
weighting can be applied even to the angular distribution of scattered
photons: it can be used to improve sampling in selected regions inside the
cloud and to lower the noise in the scattered flux observed in the selected
general direction
\item weighting schemes are easy to implement and do not significantly affect
the run times per photon package: they should be used in most calculations
\item use of quasi random numbers (instead of pseudo random numbers) brought
some improvement, but only in
the case of pure forward scattering did the the convergence approach $1/n_{\rm p}$
\item the use of a reference field improves efficiency of 
calculations when several iterations are needed: emission from constant
radiation sources (e.g., background) needs to be simulated only once and the
number of other photon packages can be reduced leading to very significant
savings in run times
\end{itemize}

The convergence (as function of the number of iterations) can be improved by
simple extrapolation that is based on dust temperature from previous
iterations or by using accelerated Monte Carlo (AMC) methods. These methods
were, for the first time, used in Monte Carlo calculations of dust emission.
Our tests showed that:

\begin{itemize}
\item heuristic extrapolation of temperature values requires some extra
storage but involves very little computations and can reduce run times by a
factor of a few
\item accelerated Monte Carlo method where internal emission-absorption- cycle
is eliminated 
is equally effective and also incurs very little computational overhead
\item for one-dimensional models one should consider including interactions
between neighbouring cells;
the run time per iteration remains essentially unchanged but the number of
iterations is significantly lower for optically thick models
\item heuristic extrapolation can be successfully combined with accelerated
Monte Carlo methods
\end{itemize}


Considering the storage and computational overhead all methods (weighted
sampling, reference field, heuristic extrapolation and accelerated Monte
Carlo) are also suitable for the use with two- and three-dimensional models. 
Tests with 1D cloud models showed that when weighted sampling and AMC methods
are used the run times are very similar to those of the Bjorkman \& Wood
(\cite{bjorkman01}) simulation scheme. However, for larger and more complex
models (e.g., with several dust species) the memory requirements are lower and
the the run times may compare favourably with the Bjorkman \& Wood method.

\begin{acknowledgements}
We thank S. Wolf for valuable comments on the BW simulation method and for his
help in the use of the MC3D program.
M.J. acknowledges the support of the Academy of Finland Grants no.
1011055, 1206049, 166322, 175068, 174854. 
\end{acknowledgements}

\appendix

\section{Implementation of the program} \label{sect:app_implementation}

\subsection{The basic program} \label{sect:app_basic}

The methods discussed in this paper are implemented in the radiative transfer
program used in Juvela \& Padoan (\cite{juvela03}). The simulated spectral
range is covered by a discrete set of frequencies. At each frequency the
radiation field and the resulting dust absorptions are simulated with a number
of photon packages. These are sent separately from discrete sources (e.g., the
central star in models $N_1$ and $N_2$), from the isotropic background and
from the dust filled cloud volume. In the basic method the true number of
photons emitted from each of these sources is divided equally between photon
packages sent from that source.

On the stellar surface and on the border of the spherically symmetric cloud
the photon packages are created at locations $(r cos\phi_i sin\theta_i, r
sin\phi_i sin\theta_i, r cos\theta_i)$, where $r$ is the radius of the object.
The random angles $\theta_i$ and $\phi_i$ are obtained from equations
\begin{equation}
    \theta_i = acos(u), \quad \phi_i = 2\pi u,
\end{equation}
where $u$ denotes a uniform random numbers in the interval $[0,1[$. For dust
emission within the cloud volume the radius $r$ is replaced with a
random variable $r_i= u^{1/3} R_{\rm cloud}$ (see below).

For emission from stellar surface and the background the photon packages
have initial directions over $2\pi$ solid angle that are determined by
random angles
\begin{equation}
        \theta_i = acos(\sqrt{u}), \quad \phi_i = 2\pi u.
\end{equation}
Here $\theta_i$ is an angle from the normal of the surface and $\phi_i$
rotation around the normal. For dust emission the packages are uniformly
distributed over $4\pi$ radians and angles $\theta$ (measured from arbitrary
reference direction) are 
\begin{equation}
        \theta_i = acos(u)
\end{equation}

Consider a step for which the optical depth of scattering is $\tau_{scat}$.
After this step the fraction of unscattered photons is $1-e^{-\tau_{scat}}$.
This is also the cumulative probability distribution for the photon free paths
and the inverse functions gives the formula for simulation of random free
paths,
\begin{equation}
    \tau_{sca}^0 = - ln(u).
\end{equation}
In this equation $u$ is again a uniformly distributed random number. When
photon package moves through a cell the optical depth of absorption,
$\tau_{\rm abs}$, is calculated and the number of absorbed photons is removed from
the photon package and added in an absorption counter in the cell. If photon
package has initially $n_{\nu}^{0}$ photons the number of absorptions is
\begin{equation}
     \Delta n_{\nu}= n_{\nu}^{0} (1-e^{-\tau_{\rm abs}}).
\end{equation}
When the total optical depth of scattering reaches $\tau_{sca}^0$ the photon
package is scattered toward a new direction that is determined by the dust
scattering function. A new value of $\tau_{sca}^0$ is generated 
and simulation continues until the number of photons remaining in the
photon packages has become insignificant or the package exits the cloud. If
package did scatter before exiting it can be registered as a sample of the
scattered flux.

In an alternative simulation scheme the free path is calculated as above,
$\tau^0 = - ln(u)$, and scattering occurs when total optical depth along the
path reaches this value.  Absorptions are not calculated along the photon
path. When package scatters the fraction of photons corresponding to the dust
albedo is retained in the package and the rest are added to the absorption
counters of the current cell. This scheme is faster when optical depth is
large but leads to poor sampling in regions of low optical depth. This method
was not used in this paper.

In some runs the method of forced first scattering (e.g., Mattila
\cite{mattila70}) was also used. In that method one first calculates the
optical depth $\Sigma \tau_{sca}$ to the cloud edge along the original
direction. The fraction $e^{-\tau_{sca}}$ of moves along this line without
scattering. The remaining fraction, $1-e^{-\tau_{sca}}$, does scatter at least
once and a conditional probability of their free path will be calculated:
'where will a photon scatter if it does scatter before the cloud border'.  The
normalized cumulative probability density function is
\begin{equation}
   P(\tau) = \frac{1- e^{-\tau_{\rm sca}}}{1-e^{-\Sigma \tau_{\rm sca}}},
\end{equation}
and random free paths are calculated using function $P^{-1}$,
\begin{equation}
   \tau_{sca}^0 = - ln ( 1 - u\,(1-e^{-\Sigma \tau_{sca}}) ).
\end{equation}
The scattered part is calculated explicitly for each package and this
improves accuracy of the estimated scattered flux, especially if the
optical depth is low.

\subsection{Weighted sampling} \label{sect:app_weighting}

In the following function $p(\vec r, \Omega)$ denotes the original probability
density function that in normal Monte Carlo runs determines the number of
photon packages sent from different position, $\vec r$, and towards different
directions, $\Omega$. The photon {\em packages} can be re-distributed
according to another probability function $q(\vec r, \Omega)$. The probability
distribution of the {\em actual photons} must not be changed and this is
accomplished by multiplying the number of true photons included in the package
by the ratio
\begin{equation}
      W = \frac{p(\vec r, \Omega)}{q(\vec r, \Omega)}.
\label{eq:weight2}      
\end{equation}
If number of packages is increased in some region ($q>p$) the weight of that
package is correspondingly decreased ($W<1$).

\subsubsection{Method $A$ - positions of external photon packages}

For external photons entering a spherical cloud the original probability
distribution of the impact parameter $d$ is $p(d)= 2 (d/R_{\rm cloud})$, where
$R_{\rm cloud}$ is the cloud radius. The cumulative probability distribution
is $P(d)=(d/R_{\rm cloud})^2$ and random impact parameters are $d_i=
P^{-1}(u)=\sqrt{u} R_{\rm cloud}$, where $u$ is a uniform random number. In
addition to $d$ one needs a random angle $\phi \in [0,2\pi]$ to determine a
rotation in the plane perpendicular to the direction of the incoming photon
package. Uniform sampling is not adequate if the innermost shells are very
small. The sampling of the inner regions could be improved, for example, by
using a centrally peaked probability distribution $q(d)= (1+\alpha) (d/R_{\rm
cloud})^{\alpha}$ with $\alpha<1$. In this case the cumulative probability
density function is $Q(d)=(d/R_{\rm cloud})^{1+\alpha}$ and the inverse
function gives formula $d_i = R_{\rm cloud} u^{(1+\alpha)^{-1}}$ ($\alpha \ne
-1$) for generating impact parameters. Usually one would include in the
package $n_{\nu}$ actual photons corresponding to the total number of photons
entering the cloud divided by the number of photon packages generated from the
background. Now this number must be scaled with $W=p(d_i)/q(d_i)$,
\begin{equation}
        n_{\nu}\prime = n_{\nu} \frac{p(d_i)}{q(d_i)} = n_{\nu} \frac{2}{1+\alpha}
	R^{\alpha-1}_{\rm cloud} d_i^{1-\alpha}.
\end{equation}
In this paper we use a scheme where identical number of photon packages is
sent towards each annulus as defined by the radial discretization. The
annulus is selected systematically and within the annulus the probability
follows the default $p(d)\sim d$ dependence,
\begin{equation}
        d_i = \sqrt{(\frac{R_{k-1}}{R_{\rm cloud}})^2 
	+ u [ (\frac{R_{k}}{R_{\rm cloud}})^2 -  
	(\frac{R_{k-1}}{R_{\rm cloud}})^2 ]}
\end{equation}
The index $k$ refers to the selected annulus and $R_k$ is outer radius of
the shell $k$. The photon numbers are scaled with the ratio between the
probability of selecting this annulus in the normal case and in the new
scheme. The first probability is proportional to the area of the annulus
and, since the annulus is selected systematically, the latter is
simply one over the number of shells, $1/N_{\rm shells}$. The weight is
\begin{equation}
        W_i = [ (\frac{R_{k}}{R_{\rm cloud}})^2 -
	       (\frac{R_{k-1}}{R_{\rm cloud}})^2 ]
	       \times N_{\rm shells}.
\end{equation}

\subsubsection{Method $B$ - positions of internal photon packages}

For emission within the cloud volume the original cumulative probability
distribution for distances from the cloud centre is $P(r)=(r/R_{\rm cloud})^3$
and usually distances would be generated from formula $r_i = P^{-1}(u) =
R_{\rm cloud}\,u^{1/3}$. We are using a scheme where shells are selected
systematically and within each shell we use the normal probability
distribution, $\sim r^3$. After selection of the shell $k$ the actual radius
for the emission event is generated from equation
\begin{equation}
    r_i  = \sqrt{ (\frac{R_{k-1}}{R_{\rm cloud}})^3 + 
    u [ (\frac{R_{k}}{R_{\rm cloud}})^3 - 
    (\frac{R_{k-1}}{R_{\rm cloud}})^3) ] }.
\end{equation}
The weight of the photon package is equal to the the cell volume (relative to
cloud volume) multiplied by the number of shells, 
\begin{equation}
        W_i = [(\frac{R_{k}}{R_{\rm cloud}})^3 -
	(\frac{R_{k-1}}{R_{\rm cloud}})^3]
	\times N_{\rm shells}.
\end{equation}

\subsubsection{Methods $C$ and $D$ - angular distribution}

Weighting can be applied to angular distributions of emitted and scattered
photons. For emitted photons the original distribution is uniform while for
the scattered photons the original distribution $p(\theta,\phi)$ is determined
by the scattering function. The scattering function is calculated in a
coordinate system fixed by the original direction of the package. In both
cases we re-distribute photon packages according to an distribution
$q(\theta)\propto e^{-\gamma \theta}$. In this equations $\theta$ is an angle
from a selected direction $\vec T$ which can point, e.g., towards the observer
or the cloud centre. After normalization this function becomes
\begin{equation}
     q(\theta) =  \frac{\gamma e^{-\gamma \theta}}{ 1- e^{- \gamma \pi}},
\end{equation}
and the cumulative probability density function is
\begin{equation}
     Q(\theta) = \frac{1-e^{-\gamma\theta}}{1-e^{-\gamma \pi}}.
\end{equation}
Function $Q^{-1}$ is used to generate random angles from distribution
$q(\theta)$,
\begin{equation}
     \theta_i = -\frac{1}{\gamma} ln[  1 - u ( 1-e^{-\gamma \pi} ) ].
\end{equation}
The rotation around the direction toward $\vec T$ is specified by a uniformly distributed 
random angle $\phi_i\in [0, 2\pi]$. In normal Monte Carlo simulation the
distribution of emitted packages is isotropic, $p(\theta)=\frac{1}{2} sin \theta$, and
the weight is obtained from Eq.~\ref{eq:weight2} as
$p(\theta)/q(\theta)$. For scattering the original angular distribution is not
uniform and we have weights
\begin{equation}
     W_i    =  \frac{p(\theta_i{^\prime}, \phi_i{^\prime})}{ q(\theta_i, \phi_i) }.
\end{equation}
The scattering function (i.e. function $p$) involves angles
$\theta_i{^\prime}$ and $\phi_i{^\prime}$ that are defined relative to the
original direction of the package while in the function $q$ the angles
$\theta_i$ and $\phi_i$ are relative to the selected direction $\vec T$.
Functions $p$ and $q$ are uniform with respect to the angles $\phi^{\prime}$
and $\phi$, respectively. However, because of the rotation between the two
coordinate systems the value of $\theta^{\prime}$ depends on the selected
angle $\phi_i$ and the weight $W$ depends the selected values of both $\theta$
and $\phi$.

\subsection{Accelerated Monte Carlo}   \label{sect:app_amc}

The basic idea behind Accelerated Monte Carlo (AMC) methods is that part of
the radiative interactions are treated explicitly when dust temperatures are
updated. The methods are completely analogous with ALI methods but
implementation differs since we compute numbers of absorbed photons and not
intensities. 

Each cell has counters for the number of photons that are absorbed in that
cell during one simulation of the radiation field. New dust temperature
estimates are calculated by balancing the absorbed energy with emission that
depends on the dust temperature. This calculation takes into account the
change in the emitted energy but not the effect that a change in the
temperature has on absorptions. In an optically thick cell the absorption
counters contain mostly photons that were originally emitted within the same
cell. If temperature is, for example, increased the calculations assume that
all added emission escape the cell. In reality most photons are re-absorbed
and the net flow of energy out from the cell is much smaller. Consequently,
the needed temperature correction is severely underestimated and convergence
remains slow.

In AMC this problem is removed by treating explicitly those photons that are
absorbed within the same cell from which they were emitted.  The simulation
proceeds in normal fashion except that absorbed photons are counted separately
depending on whether they were originally emitted from the same cell or not.
We denote these counters with $n^{\rm abs}_{\rm int}$ and $n^{\rm abs}_{\rm
ext}$ the indices referring to the origin of the photons (internal vs.
external). In normal runs the absorption counters (number of photons per unit
frequency interval) correspond to the sum of these two counters and the dust
temperatures $T_{\rm d}$ is solved from an equilibrium condition for unit
volume,
\begin{equation}
   \int  n^{\rm abs}(\nu) \, h \nu \, d\nu = 
   4\pi \int \kappa_{\nu} B_{\nu}(T_{\rm d}) d\nu.
\end{equation}
In AMC schemes this equation will be modified. In addition to $n^{\rm
abs}_{\rm int}$ we have also information about the number photons emitted
from the cell, $n^{\rm emit}$, and we can calculate photon escape
probabilities
\begin{equation}
    \beta =   \frac{ n^{\rm emit} - n^{\rm abs}_{\rm int}}{n^{\rm emit}}.
\end{equation}
The equilibrium equation can be re-written in the form
\begin{equation}
   \int  ( n^{\rm abs}_{\rm int} + n^{\rm abs}_{\rm ext} ) \ h\nu  d\nu =
   4\pi \int \kappa_{\nu} B_{\nu}(T_{\rm d}) [ (1-\beta) + \beta ] d\nu.
\end{equation}
On the right hand side the part $1-\beta$ represents photon emitted and
absorbed in the same cell. On the left side this corresponds to counters
$n^{\rm abs}_{\rm int}$. This part can be subtracted and we are left with a
condition
\begin{equation}
   \int  n^{\rm abs}_{\rm ext} \, h \nu  d\nu =
      4\pi \int \kappa_{\nu} B_{\nu}(T_{\rm d}) \beta d\nu.
\label{eq:equilibrium_diag}      
\end{equation}
The left hand side is the net inflow of energy into the cell, the right hand
side is the net energy loss and the cycle of local absorptions and emissions
has been eliminated. This form will lead to faster convergence, especially if
escape probability $\beta$ was small. In ALI terms this corresponds to a
separation of the diagonal part of the lambda operator. The lambda operator is
defined by the relation $J=\Lambda S$ and diagonal elements describe the
effect of the local source function on the local mean intensity. In AMC this
separation was accomplished by using separate counters, $n^{\rm abs}_{\rm
int}$. In our implementation the temperature is solved from
Eq.~\ref{eq:equilibrium_diag} with Newton-Rhapson iteration. In large models
the run times could be decreased by pre-calculating a direct mapping between
incoming energy (left side of Eq.~\ref{eq:equilibrium_diag}) and dust
temperature. That way the solution of Eq.~\ref{eq:equilibrium_diag} would
involve only a single table look-up (and possibly an interpolation). However,
Eq.~\ref{eq:equilibrium_diag}) represents a small fraction of all computations
and the effect on run times would be negligible.

The AMC methods can be extended by considering radiative interactions between
cells. Additional counters can be used to register separately those of the
absorbed photons that were emitted from an immediate neighbour. In a spherical
model this means the neighbouring shells with indices $k-1$ and $k+1$. The
total number of absorbed photons can be written
\begin{equation}
n^{\rm abs}_{k} = n^{\rm abs}_{k,\,\rm int} + n^{\rm abs}_{k,\,\rm ext} + 
n^{\rm abs}_{k-1,k} +  n^{\rm abs}_{k+1,k},
\end{equation}
where terms $n^{\rm abs}_{i,j}$ correspond to photons emitted in cell $i$ and
absorbed in cell $j$. This time the counter $n^{\rm abs}_{\rm ext}$ includes
only photons originating beyond the neighbouring cells. Together with the
number of emitted photons these define ratios
\begin{equation}
\xi_{i,j}  =  \frac{n^{\rm abs}_{i,j}}{n^{\rm emit}_{i}},
\end{equation}
for $i=j\pm 1$. One can also define $\xi_{k,k}$ which is equal to $1-\beta$. The terms
$\xi_{i,j}$ describe the radiative interaction between cells: what part of
photons emitted in cell $i$ is absorbed in cell $j$. Actually, $\xi_{i,j}$ is
based on photon numbers per unit volumes so that in the following the cell
volumes are not explicitly visible. In the equilibrium equation counters
$n^{\rm abs}_{k\pm 1,k}$ can be replaced with calculated emission from shell
$k\pm1$ multiplied with the corresponding factor $\xi_{k\pm1,k}$. The escape
probabilities $\beta$ are again used to eliminate local absorptions, and the
equilibrium condition for cell $k$ becomes
\begin{eqnarray}
   \int  n^{\rm abs}_{k,\,\rm ext}(\nu) \ h\nu \, d\nu \nonumber \\
   +  4\pi \int \kappa_{k-1}(\nu) B_{\nu}(T_{\rm d}^{k-1}) \, \xi_{k-1,k}(\nu) \, d\nu 
   \nonumber \\
   +  4\pi \int \kappa_{k+1}(\nu) B_{\nu}(T_{\rm d}^{k+1}) \, \xi_{k+1,k}(\nu) \, d\nu
   \nonumber \\
   =    4\pi \int \kappa_{k}(\nu) B_{\nu}(T_{\rm d}^{k}) \, \beta(\nu)  d\nu
\label{eq:equilibrium_triag}   
\end{eqnarray}
The right hand side identical with Eq.~\ref{eq:equilibrium_diag}. The left
hand side depends now explicitly on the dust temperatures of the neighbouring
cells and all temperatures must be solved from a nonlinear set of equations.
This has the advantage that temperature changes in neighbouring cells are
taken into account immediately and not only after next simulation step. If
shells are optically thick there will be little radiation coming from beyond
the neighbouring shells ($n^{\rm abs}_{k,\,\rm ext}\approx 0$) and the
solution of the Eq.~\ref{eq:equilibrium_triag} is immediately close to the
true solution. In ALI terms this corresponds (in one-dimensional case) to the
use of a tri-diagonal operator. In our implementation Newton iterations are
used to solve for one temperature while other temperature values were kept
constant. This is repeated for all cells until solution is fully converged.
The procedure is not particularly efficient but still fast enough so that
overall run times per iteration were practically the same as when no AMC
methods were used.

Since coupling between all cells is not stored the simulation of the radiation
field must be repeated for each iteration in the usual fashion. The photon
escape probabilities and the coupling between neighbours remain, in principle,
unchanged. We repeat, however, also these steps so that the calculations
remain consistent if a different set of random numbers is used on different
iterations. 

Further AMC methods can be constructed by considering interactions between
more cells. For example, in a three-dimensional cartesian grid one might
consider interactions between a cell and its six neighbours along coordinate
axes. However, each interaction requires separate counters $n^{\rm abs}_{i,j}$
and the total memory requirements of the program would be increased by a
factor of a few.

\subsection{Reference field}   \label{sect:app_reference}

Normally Monte Carlo sampling is used separately on each iteration to estimate
the strength of the total radiation field. If emissions and absorptions are
known for some reference situation, we need to estimate only the difference
between the true and the reference field and sampling errors are
correspondingly smaller.  Normal Monte Carlo calculations are based on the
current estimate of the number of photons emitted from each cell, $n^{\rm
emit}$, and the number of photons absorbed in each cell, $n^{\rm abs}$, which
results from the simulation of the radiation field. We use as the reference
the solution from the previous iteration. Before dust temperatures are updated
the numbers are are copied as counters for the reference field
\begin{equation}
n^{\rm abs}_{\rm ref} \,  \leftarrow \, n^{\rm abs}, \quad
n^{\rm emit}_{\rm ref} \, \leftarrow \, n^{\rm emit}.
\end{equation}
As new dust temperatures are solved we get new estimates for $n^{\rm emit}$
and the following simulation is used to determine the change from the current
reference situation i.e. from the previous iteration. Therefore, simulated
photon packages correspond to differences
\begin{equation}
  \Delta n^{\rm emit} =  n^{\rm emit} - n^{\rm emit}_{\rm ref},
\end{equation}
and the resulting absorptions, $\Delta n^{\rm abs}$, constitute a correction
to the number of absorptions found in the reference case. The closer the
reference field is to the actual field the smaller are the $\Delta n^{\rm
emit}$ and $\Delta n^{\rm abs}$ terms and the smaller the noise in the
estimated difference between the `true' field (corresponding to the current
temperature estimates) and the reference field. Note that $\Delta n^{\rm
emit}$ and $\Delta n^{\rm abs}$ can also be negative. After the simulation new
temperatures are calculated using the sum of the reference field and the
corrections 
\begin{equation}
   n^{\rm abs} =  n^{\rm abs}_{\rm ref} + \Delta n^{\rm abs},
\label{eq:ref1}   
\end{equation}
and at the same time, before new values for $n^{\rm emit}$ are calculated, the
description of the reference field is updated
\begin{equation}
n^{\rm abs}_{\rm ref}  \, \leftarrow  n^{\rm abs}_{\rm ref} + \Delta
n^{\rm abs}, \quad 
n^{\rm emit}_{\rm ref} \, \leftarrow   n^{\rm emit}.
\label{eq:ref2}   
\end{equation}
As the reference get closer to the true solution also the fluctuations from
the simulation of the difference field become smaller and, in principle, the
final noise depends on the total number of photon packages and not just on the
number of packages simulated during one iteration. Therefore, given noise
level can be obtained using a lower number of photon packages per iteration.
Another advantage is that emission from background and discrete sources need
to be simulated only once. On the following iterations their effect is already
included in the reference field.

The previous scheme is not optimal in the sense that iterations before large
temperature corrections have larger impact on the final solution. In an
extreme case, if solution converges after one iteration the current
fluctuations are 'frozen' in the solution which does not change on any of the
following iterations. This problem can be avoided by using as the reference a
running average of the $n^{\rm abs}$ and $n^{\rm emit}$ where early iterations
(when the solution is still far from the correct one) are given a smaller
weight. Alternatively, one can divide the emission of the photons from heating
sources over many iterations thus avoiding a large temperature jump on early
iterations. As the solution converges at a nearly uniform pace all iterations
contribute equally to the estimated reference field and the final noise level
is correspondingly lower. In practise, this method was found to work very well.
It does slow down the convergence but the impact is probably no more than
$\sim$50\%. The method was used in Sect.~\ref{sect:comparison} where emission
of the photons from the central source was divided equally over $\sim$2/3 of
the total number of iterations.

A reference field can be used together with AMC methods. As long as optical
depths are temperature independent each iteration gives additional samples for
the estimation of escape probabilities and the strength of the radiative
coupling between cells. Equations~\ref{eq:ref1}--\ref{eq:ref2} can be applied
to counters $n^{\rm abs}_{\rm int}$, $n^{\rm abs}_{\rm ext}$ etc. or one can
directly calculate a suitably weighted running averages of the $\beta$ and
$\xi_{i,j}$ factors.

\end{document}